\documentclass[preprint,11pt]{aastex}
\newcommand{\kms}{km s$^{-1}$}
\newcommand{\zabs}{$z_{\rm abs}$}
\newcommand{\lya}{Ly$\alpha$\ }

\newcommand{\nav}{$N_{\rm a}(v)$}

\slugcomment{Accepted for publication in {\it The Astrophysical Journal}}
\shortauthors{TRIPP et al.}
\shorttitle{Damped Absorber in Virgo}
\begin{document}

\title{Discovery of a Primitive Damped Lyman $\alpha$ Absorber Near an
X-ray Bright Galaxy Group in the Virgo Cluster\altaffilmark{1}}

\altaffiltext{1}{Based on observations with (1) the NASA/ESA {\it
Hubble Space Telescope}, obtained at the Space Telescope Science
Institute, which is operated by the Association of Universities for
Research in Astronomy, Inc., under NASA contract NAS 5-26555, and (2)
the NASA-CNES/ESA {\it Far Ultraoviolet Spectroscopic Explorer}
mission, operated by Johns Hopkins University, supported by NASA
contract NAS 5-32985.}

\author{Todd M. Tripp,\altaffilmark{2} Edward
B. Jenkins,\altaffilmark{3} David V. Bowen,\altaffilmark{3} Jason
X. Prochaska,\altaffilmark{4} Bastien Aracil,\altaffilmark{2} and Rajib
Ganguly\altaffilmark{5}}

\altaffiltext{2}{Department of Astronomy, University of Massachusetts,
Amherst, MA 01003, Electronic mail: tripp@astro.umass.edu}

\altaffiltext{3}{Princeton University Observatory,
Peyton Hall, Princeton, NJ 08544}

\altaffiltext{4}{University of California Observatories, Natural
Sciences II Annex, UC-Santa Cruz, Santa Cruz, CA 95064}

\altaffiltext{5}{Space Telescope Science Institute, 3700 San Martin
Drive, Baltimore, MD 21218}

\begin{abstract}
We present a new ultraviolet echelle spectrum of PG1216+069, obtained
with {\it HST}+STIS, which reveals damped Lya (DLA) absorption as well
as \ion{O}{1}, \ion{C}{2}, \ion{Si}{2}, and \ion{Fe}{2} absorption
lines at \zabs\ = 0.00632 near the NGC4261 galaxy group in the
periphery of the Virgo cluster. The absorber shows no evidence of
highly-ionized gas, which places constraints on ``warm-hot'' missing
baryons in the X-ray bright NGC4261 group. The well-developed damping
wings of the \lya line tightly constrain the \ion{H}{1} column
density; we find log N(\ion{H}{1}) = 19.32$\pm$0.03. The metallicity
of this sub-DLA is remarkably low, [O/H] = $-1.60^{+0.09}_{-0.11}$,
which is comparable to many analogous high-redshift systems, and the
iron abundance indicates that this absorber contains little or no
dust. Nitrogen is underabundant: we detect neither \ion{N}{1} nor
\ion{N}{2}, and we show that the absence of nitrogen is not due to
ionization effects but rather indicates that [N/O] $\leq -0.28 \
(3\sigma)$. Despite the proximity of the sight line to NGC4261 group,
there are no bright galaxies with small impact parameters at the
absorption redshift: the nearest known galaxy is a sub$-L*$ galaxy
with a projected distance $\rho = 86 h_{75}^{-1}$ kpc, while the
closest $L*$ galaxy is NGC4260 at $\rho = 246 h_{75}^{-1}$ kpc. The
low metallicity and nitrogen underabundance indicate that this low$-z$
sub-DLA is a relatively primitive gas cloud. We consider the nature
and origin of the sub-DLA, and we discuss several possibilities. The
properties of the sub-DLA are similar to those of the interstellar
media in blue compact dwarf galaxies and are also reminiscent of Milky
Way HVCs. The sub-DLA could also be related to a dwarf spheroidal
galaxy, if the absorption arises in gas ejected or stripped from such
an object. Finally, the object could simply be a small dark-matter
halo, self-enriched by a small amount of internal star formation but
mostly undisturbed since its initial formation.  In this case, the
small halo would likely be an ancient building block of galaxy
formation that formed before the epoch of reionization.
\end{abstract}


\section{Introduction}

Understanding the metal enrichment history of the universe, in
environments ranging from the low-density intergalactic medium to the
highest overdensity regions in galaxies and galaxy clusters, has
emerged as one of the pressing challenges of current cosmology and
galaxy evolution studies. Measurements of abundances in QSO absorption
line systems have shown remarkably little evolution of metallicity
from \zabs\ $>$ 4 down to \zabs\ $\approx$ 1. Moreover, this lack of
evolution is evident in all absorber samples regardless of \ion{H}{1}
column density, ranging from the low-density \lya clouds with
$N$(\ion{H}{1}) $< 10^{14}$ cm$^{-2}$ (e.g., Songaila 2001; Pettini et
al. 2003; Schaye et al. 2003; Simcoe, Sargent, \& Rauch 2004) up to
the highest-column density damped \lya absorbers with $N$(\ion{H}{1})
$> 10^{20}$ cm$^{-2}$ (e.g., Pettini et al. 1999; Prochaska \& Wolfe
2002); these absorbers span several orders of magnitude in overdensity
$\delta \ (\equiv \rho/<\rho >)$. It does appear that at any
particular redshift, the lower-$N$(\ion{H}{1})/lower$-\delta$
absorbers generally have lower overall metallicities, but it should be
noted that as $N$(\ion{H}{1}) decreases, the abundance measurements
become increasingly sensitive to ionization corrections (see, e.g.,
Figure 13 in Schaye et al. 2003). There are also indications that
absorber metallicities are highly inhomogeneous (Schaye et al. 2003),
and it remains possible that some absorption systems are metal-free
(e.g., Levshakov et al. 2003; Simcoe et al. 2004; Aracil et al. 2004). 

For the damped \lya absorbers (DLAs) in particular, the observation of
weak evolution with redshift is intriguing. After all, the Milky Way
is a damped \lya absorber, and its ISM metallicity (in the vicinity of
the Sun) is substantially higher than the DLA metallicity trend
extrapolated to $z =$ 0. If the DLAs arise in the bound interstellar
media of objects that are the precursors of today's spiral galaxies,
which has long been the canonical view (e.g., Wolfe et al. 1986), then
why are the abundances not increasing at a rate sufficient to reach
solar metallicity at $z \approx$ 0? Recently, Prochaska et al. (2003)
have presented evidence of slowly increasing metallicity with
decreasing $z$ based on a sample of 125 DLAs, but they note that the
extrapolation to $z$ = 0 still falls $4\sigma$ below the Milky Way
metallicity. We note, though, that the evolution of DLA metallicity
with redshift still suffers from considerable uncertainty due to the
limited number of robust abundance measurements in low$-z$ DLAs (see
below). Theoretical studies employing hydrodynamic simulations have
reproduced the slow evolution of the DLAs but generally predict
substantially higher mean metallicities than observed (Cen et
al. 2003; Nagamine, Springel, \& Hernquist 2004), a discrepancy that
is attributed to either a bias in the observations due to obscured
QSOs (but see counterarguments in Ellison et al. 2001; Prochaska 2003)
or inadequate treatment of supernova feedback and the multiphase
nature of the gas.

However, all current observational studies are dominated by absorbers
with \zabs\ $\gtrsim$ 2. The lookback time increases quickly with
redshift, and the current samples only probe the first $\sim$20\% of
the history of the universe. Indeed, if the DLA metallicities reported
by Prochaska et al. (2003) are expressed as a function of cosmic time
instead of redshift, then these latest data indicate that DLA
metallicites appear to be doubling every billion years at $z \approx
3$.  Without the lowest-redshift bin, extrapolation of this trend
vs. time to the current epoch would predict highly super-solar
abundances in DLAs at $z \approx$ 0. Clearly, the lowest-redshift DLAs
now present the key to understanding DLA metallicity evolution, and it
is crucial to examine how the absorption systems evolve in the
relatively unstudied 80\% of the age of the universe between $z
\approx$ 2 and now. For the most part, this requires observations from
space (the optimal absorption lines for abundance measurements
are in the UV at these redshifts), a task which presents several
challenges. Low-redshift high-$N$(\ion{H}{1}) absorption systems are
rarely found in traditionally constructed unbiased surveys; e.g.,
Jannuzi et al. (1998) found only one DLA in a survey of 83 QSOs with
the {\it HST} Faint Object Spectrograph. Rao \& Turnshek (2000) have
demonstrated that the low$-z$ DLA sample can be increased more rapidly
by selecting targets known to show the \ion{Mg}{2} $\lambda
\lambda$2796.35, 2803.53 doublet a priori, but even DLAs found in this
way are slowly yielding metallicities, partly because the background
QSOs are still too faint for follow-up high-resolution spectroscopy
with current space-borne instrumentation.  The lowest-redshift DLAs
are especially interesting because these absorbers provide the
greatest opportunity to understand the nature of the DLA
progenitor. For example, we have found a DLA that originates in a
low-surface brightness galaxy at $z$ = 0.009 (Bowen et al. 2001a,b);
this galaxy would be challenging to detect at even moderate redshifts.

\begin{figure}
\plotone{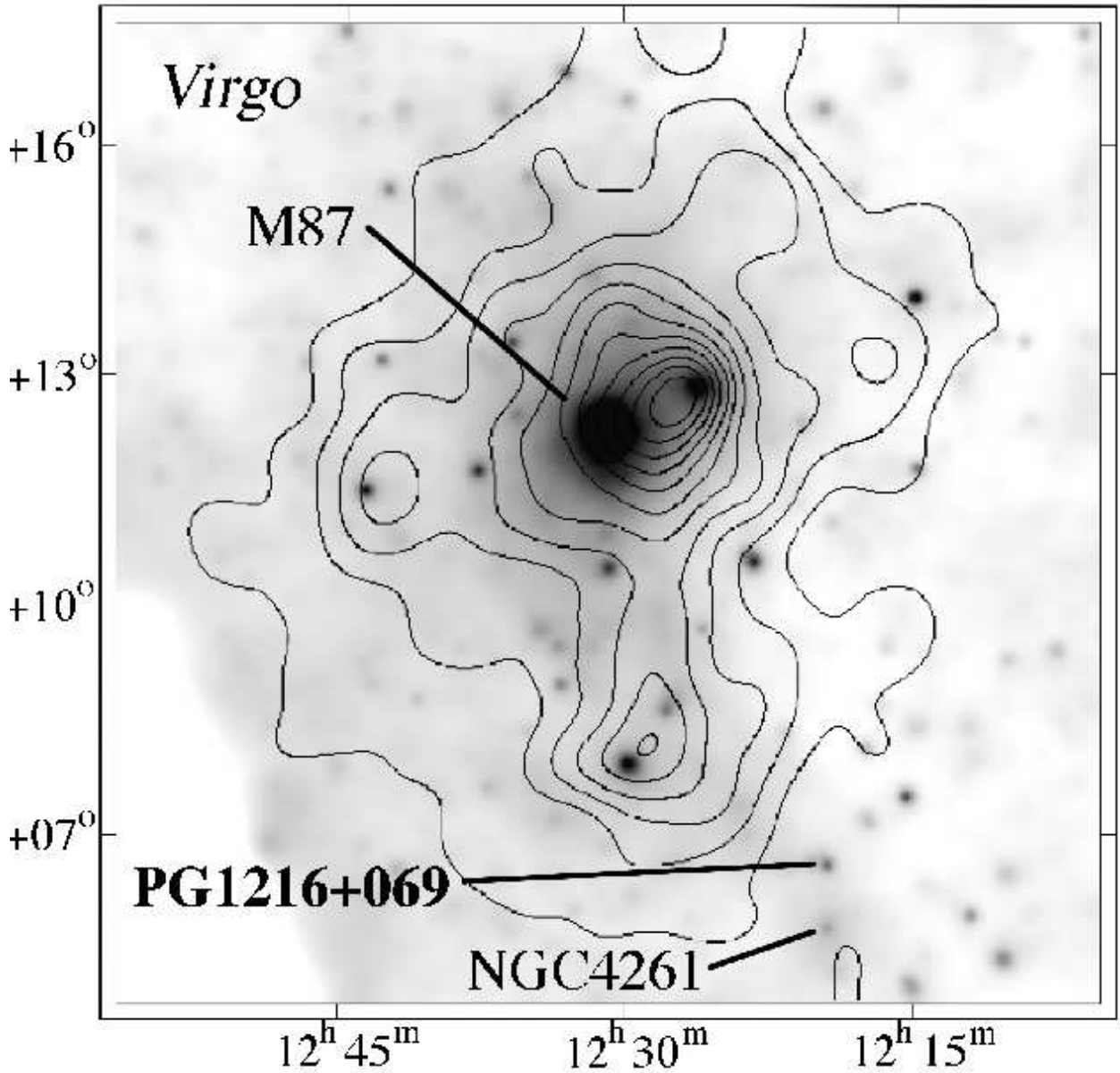}
\caption{Location of the sight line to PG1216+069 plotted on the
Schindler et al. (1999) map of X-ray emission in the vicinity of the
Virgo cluster from the ROSAT all-sky survey (logarithmic grayscale;
see also B\"{o}hringer et al. 1994) and galaxy number density from the
VCC catalog (contours; see also Binggeli et al. 1993). The X-ray and
galaxy number density maps have both been smoothed with a Gaussian
with $\sigma$ = 24'. The locations of M87 and the X-ray-bright
NGC4261 group are also marked; see Davis et al. (1995) for a deeper,
pointed ROSAT observation of the NGC4261 group.\label{xraymap}}
\end{figure}

For these reasons, we are pleased to report in this paper the
serendipitous discovery of a new damped \lya system at \zabs\ =
0.00632 in the direction of PG1216+069.  Using the E140M echelle mode
of the Space Telescope Imaging Spectrograph (STIS) on board {\it HST},
we have been conducting a large survey for low$-z$ \ion{O}{6}
absorption lines, which appear to harbor a significant fraction of the
baryons at the present epoch (Tripp et al. 2000a,b; Savage et
al. 2002; Richter et al. 2004; Sembach et al. 2004a). The sight line to
one of the QSOs observed under the auspices of this \ion{O}{6} survey,
PG1216+069, passes through a very interesting region of the nearby
universe. As shown in Figure~\ref{xraymap}, this sight line pierces
the southwest periphery of the Virgo galaxy cluster. The PG1216+069
sight line is outside of the 6$^{\circ}$ cluster core defined by Tully
\& Shaya (1984), but it is within the sphere of influence of the
cluster according to those authors. An even more intriguing aspect of
this sight line is that it is $\sim$400 kpc in projection from the
center of the NGC4261 galaxy group. This group, which is at a redshift
close to (but slightly higher than) the redshift of the Virgo cluster
proper, is known to have a hot intragroup medium from {\it ROSAT}
detection of diffuse X-ray emission (Davis et al. 1995).

While observations of PG1216+069 made with the first-generation {\it
HST} spectrographs had revealed a strong \lya line at \zabs\ $\approx$
0.006, i.e., near the redshift of the Virgo cluster (Bowen et
al. 1996; Impey, Petry, \& Flint 1999), the low spectral resolution of
the early {\it HST} spectra precluded recognition of the damped nature
of this absorption system. A portion of the new STIS spectrum of this
QSO covering the \lya line in Virgo is shown in
Figure~\ref{dlaspec}. This \lya line has well-developed damping
wings. Technically, even though the damping wings are quite obvious,
the \ion{H}{1} column density of the \zabs\ = 0.00632 system is too
low to qualify as a ``damped'' \lya absorber according to the standard
definition, which requires log $N$(\ion{H}{1}) $\geq$ 20.3.  Hence, we
will refer to this system as a ``sub-DLA''. Originally, the standard
DLA definition was mainly an observational convenience for early
low-resolution spectroscopic studies. In many respects, sub-DLAs can
have physical characteristics that are similar to the higher
$N$(\ion{H}{1}) DLAs. However, ionization corrections can be more
important in sub-DLAs (e.g., Dessauges-Zavadsky et al. 2003;
P\'{e}roux et al. 2003).

In this paper we present the remarkable properties of this new, nearby
sub-DLA in the immediate vicinity of the Virgo cluster. The manuscript
is organized as follows: after presenting the observations and
absorption line measurements in \S~\ref{obsec} and \S~\ref{abmeas},
respectively, we derive the absolute and relative metal abundances of
the absorber in \S~\ref{abund}, including analyses of potentially
important sources of systematic errors such as unresolved line
saturation and ionization corrections. In \S~\ref{physcon}, we briefly
discuss the physical conditions in the absorbing gas. We then provide
information on the environment in which the system is found
(\S~\ref{abenv}) and limits on \ion{O}{6} absorption associated with
the NGC4261 group (\S~\ref{whimsec}) before offering some comments on
the nature and implications of the absorber properies in
\S~\ref{disec}. We close with a summary of the main results in
\S~\ref{sumsec}, and the appendix briefly discusses STIS wavelength
calibration errors.

\section{Observations\label{obsec}}

PG1216+069 was observed with the E140M spectroscopic mode of STIS on 7
occasions in 2003 May-June; the total integration time was 69.8 ksec.
This echelle spectrograph provides a resolution of 7 \kms\ (FWHM) with
$\sim$2 pixels per resolution element and covers the $1150 - 1700$
\AA\ range with only a few small gaps between orders at $\lambda >
1630$ \AA\ (Woodgate et al. 1998; Kimble et al. 1998). The
observations employed the $0\farcs 2 \times 0\farcs 06$ slit in order
to minimize the wings in the line-spread function (see Figure 13.91 in
the STIS Instrument Handbook, Kim Quijano et al. 2003).  The data were
reduced as described in Tripp et al. (2001) using the STIS Team
version of CALSTIS at the Goddard Space Flight Center.

We encountered three minor calibration issues when we reduced these
data.  First, there are a significant number of warm/hot pixels on the
detector. Some of these were successfully fixed with an automatic
identification and interpolation algorithm using pixels adjacent to
the hot pixels, but many of the warm pixels are not easily identified
and corrected in the individual exposures due to low count rates, and
these remained in the data after the hot-pixel cleaning procedure.
When the individual exposures were combined into a single final
spectrum, these pixels became more evident. Fortunately, there is no
clear evidence that the important absorption lines discussed in this
paper are significantly affected by warm/hot pixels.  Second, after
flux calibration, overlapping regions of adjacent orders showed slight
flux discrepancies, mainly very close to order edges.  We used the
recently improved ripple correction (as of 2003 July), but
nevertheless these small discrepanices are likely due to residual
imperfections in the ripple correction. To avoid spurious features at
order edges, when we coadded the overlapping regions of adjacent
orders, we smoothly decreased the weight given to the 50 pixels
approaching the order edges with the last 10 pixels thrown out all
together.  Third, we noticed and corrected small errors ($\lesssim 1$
pixel) in the relative wavelength calibration.  These wavelength
calibration errors are larger than expected, and indeed are larger
than we have seen in many other observations with this mode of STIS,
so we have summarized the evidence of this problem in the
appendix. Nevertheless, compared to archival spectra available from
other UV instruments/facilities, the STIS relative wavelength
calibration is still excellent, and moreover these small shifts would
have very little effect on our measurements and science conclusions
even if they were neglected.

PG1216+069 was also observed with the {\it Far Ultraviolet
Spectroscopic Explorer (FUSE)} in a single visit on 2001 February 6
for 12.4 ksec (dataset root P10721).\footnote{For information on the
design and performance of {\it FUSE}, see \S 3 in Moos et al. (2002)
and references therein.} This is a short exposure compared to typical
{\it FUSE} observations of extragalactic objects, and indeed the final
{\it FUSE} spectrum of PG1216+069 is relatively noisy.  Nevertheless,
we shall see that by virtue of its unique wavelength coverage, this
{\it FUSE} spectrum provides useful constraints for our analysis of
the Virgo sub-DLA. To reduce the {\it FUSE} data, we first calibrated
the raw time-tagged data frames using the CALFUSE pipeline reduction
(version 2.4.0). The calibrated datasets comprised six individual
sub-exposures for each {\it FUSE} channel. The signal-to-noise of each
subexposure was too low to permit us to derive any shifts between
sub-exposures, so we simply summed the total number of counts in each
pixel from all six sub-exposures.  However, our experience with {\it
FUSE} observations of substantially brighter targets indicates that
shifts between sub-exposures are small or negligible in a single
visit, so our procedure for coaddition of the PG1216+069 data should
result in little or no degradation of spectral resolution.  We binned
the final coadded {\it FUSE} spectrum to $\sim 10$ \kms\ pixels, which
provides $\sim$2 pixels per resolution element. Only the LiF channels
contained enough signal to be useful, so the effective wavelength
coverage of the {\it FUSE} spectrum is 1000-1185 \AA . The usual
strong ISM lines (e.g., \ion{Si}{2} $\lambda$1020.70, \ion{O}{6}
$\lambda \lambda$1031.93,1037.62, \ion{C}{2} $\lambda$1036.34,
\ion{Fe}{2} $\lambda$1144.94) are readily apparent in the final {\it
FUSE} spectrum along with several extragalactic absorption lines. In
this paper, we will make use of the {\it FUSE} data for two
purposes. First, we will use the LiF2a segment to search for
\ion{N}{2} $\lambda$1083.99 absorption at the redshift of the Virgo
sub-DLA. Second, we will use the LiF1a and LiF2b segments to place
constraints on \ion{O}{6} associated with the NGC4261 group and the
sub-DLA. To calibrate the zero point of the LiF2a wavelength scale, we
aligned the \ion{Fe}{2} $\lambda$1144.94 absorption profile with the
profile of the \ion{Fe}{2} $\lambda$1608.45 line in the STIS
spectrum. For the LiF1a and LiF2b spectra, we compared the \ion{C}{2}
$\lambda$1036.34 and \ion{C}{2} $\lambda$1334.53 lines from the {\it
FUSE} and STIS spectra, respectively. These \ion{Fe}{2} and \ion{C}{2}
transitions have similar strengths,\footnote{The \ion{Fe}{2} and
\ion{C}{2} lines are strongly saturated and black in their cores, but
since the profile edges are sharp and the lines have similar strengths
and equivalent widths, the STIS and FUSE data can be aligned with good
precision after the STIS resolution has been degraded to match the
FUSE resolution.} and this procedure results in a zero-point velocity
uncertainty of 5$-$10 \kms . Due to the low S/N ratios of the
individual spectra, we aligned and coadded the LiF1a and LiF2b
segments for the \ion{O}{6} measurements.  This may degrade the
spectral resolution, but examination of the coadded spectrum indicates
that the degradation is minimal. In the case of the \ion{N}{2} line,
only the LiF2a spectrum had the right wavelength coverage and
sufficient S/N to place useful constraints.

\section{Absorption Line Measurements\label{abmeas}}

\begin{figure}
\plotone{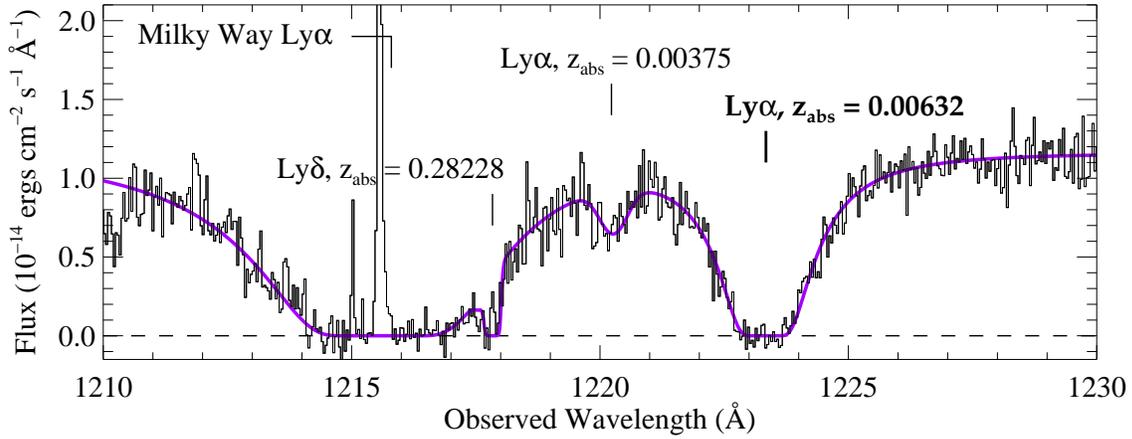}
\caption[]{Region of the STIS E140M echelle spectrum of PG1216+069
showing the sub-damped \lya absorber in the outskirts of the Virgo
galaxy cluster at \zabs\ = 0.00632. The damped \ion{H}{1} \lya line
due to the Milky Way ISM, a \lya line at \zabs\ = 0.00375, and the
Ly$\delta$ line at \zabs\ = 0.28228 are also marked.  A fit to all of
these features is overplotted with a thick line. The emission line in
the core of the Milky Way \lya profile is the geocoronal \ion{H}{1}
emission.  At full (unbinned) resolution, the STIS E140M echelle mode
provides 7 \kms\ resolution (FWHM) with $\sim$2 pixels per resolution
element. In this figure for display purposes only, the data have been
binned 3 pixels into 1. For measurements and in all other figures, we
use the full-resolution, unbinned STIS data.\label{dlaspec}}
\end{figure}

A selected portion of the final, coadded STIS spectrum is shown in
Figure~\ref{dlaspec}. The strong \lya absorption line at \zabs\ =
0.00632 is readily apparent along with adjacent lines from various
redshifts including the damped \lya line due to Galactic \ion{H}{1},
Ly$\delta$ at \zabs\ = 0.28228, and \lya at \zabs\ = 0.00375. The
Ly$\delta$ line is recorded at low signal-to-noise ratio (S/N) due to
its location in the pit of the Milky Way \lya
absorption. Nevertheless, the Ly$\delta$ identification is secure
because $>$11 other Lyman series lines are detected at \zabs\ = 0.28228.
Only \lya is detected at \zabs\ = 0.00375. We note that \lya absorbers
are frequently detected at 0.003 $\lesssim$ \zabs\ $\lesssim$ 0.008 in
the spectra of QSOs in the general direction of the Virgo cluster
(e.g., Bowen et al. 1996; Impey, Petry, \& Flint 1999; Tripp et
al. 2002; Rosenberg et al. 2003).

The well-developed Lorentzian wings of the \lya line at \zabs\ =
0.00632 tightly constrain the \ion{H}{1} column density. Since the
\zabs\ = 0.00632 \lya line is positioned in the red wing of the Milky
Way damped \lya profile (see Figure~\ref{dlaspec}), we fitted the
Milky Way \ion{H}{1} line as well as the \zabs\ = 0.00632 absorber. We
used the method and software described in Jenkins et al. (1999) to fit
the damped \lya profiles and measure $N$(\ion{H}{1}).\footnote{As
  described in Jenkins et al. (1999) and Sonneborn et al. (2000), our
  \ion{H}{1} measurement software allows the height and shape of the
  continuum to vary during the fitting process, and the error bars
  reflect the continuum placement uncertainty accordingly. Inspection
  of the black cores of strongly saturated lines in STIS echelle
  spectra occasionally reveals small errors in the STIS flux zero
  point (usually $\lesssim$ 2\% of the continuum level). Consequently,
  the flux zero level was also allowed to vary as a free parameter
  during the fitting process, and the saturated cores of the lines
  were included in the $\chi ^{2}$ calculation.} The resulting fit is
shown with a solid line in Figure~\ref{dlaspec}, and the absorber
\ion{H}{1} column densities and uncertainties are summarized in
Table~\ref{h1col}, including the weak \lya line at \zabs\ = 0.00375.
The \ion{H}{1} column derived from the Milky Way \lya profile is in
good agreement with the $N$(\ion{H}{1}) reported by Lockman \& Savage
(1995) based on 21cm emission in the direction of PG1216+069.

\begin{figure}
\epsscale{0.7}
\plotone{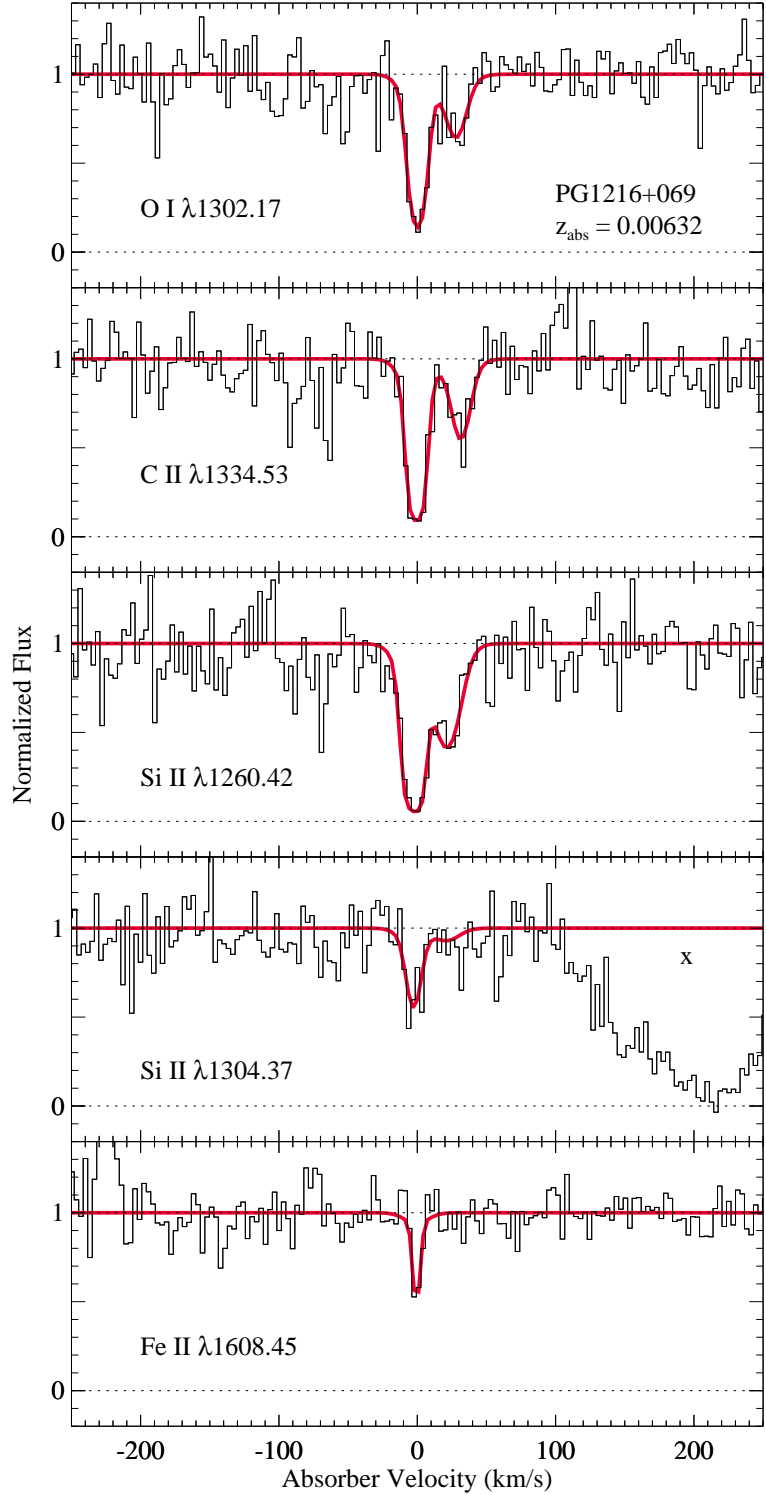}
\caption[]{\footnotesize Continuum-normalized absorption profiles of
lines detected in the sub-damped \lya absorber at \zabs\ = 0.00632
(plotted vs. velocity in the absorber frame, i.e., $v = 0$ \kms\ at
\zabs\ = 0.00632). The absorption-line fits summarized in
Table~\ref{voigtfit} are overplotted with thick
lines. An unrelated absorption feature in the \ion{Si}{2}
$\lambda$1304.37 panel is marked with an 'x'.\label{normprofs}}
\end{figure}

\begin{figure}
\epsscale{0.7}
\plotone{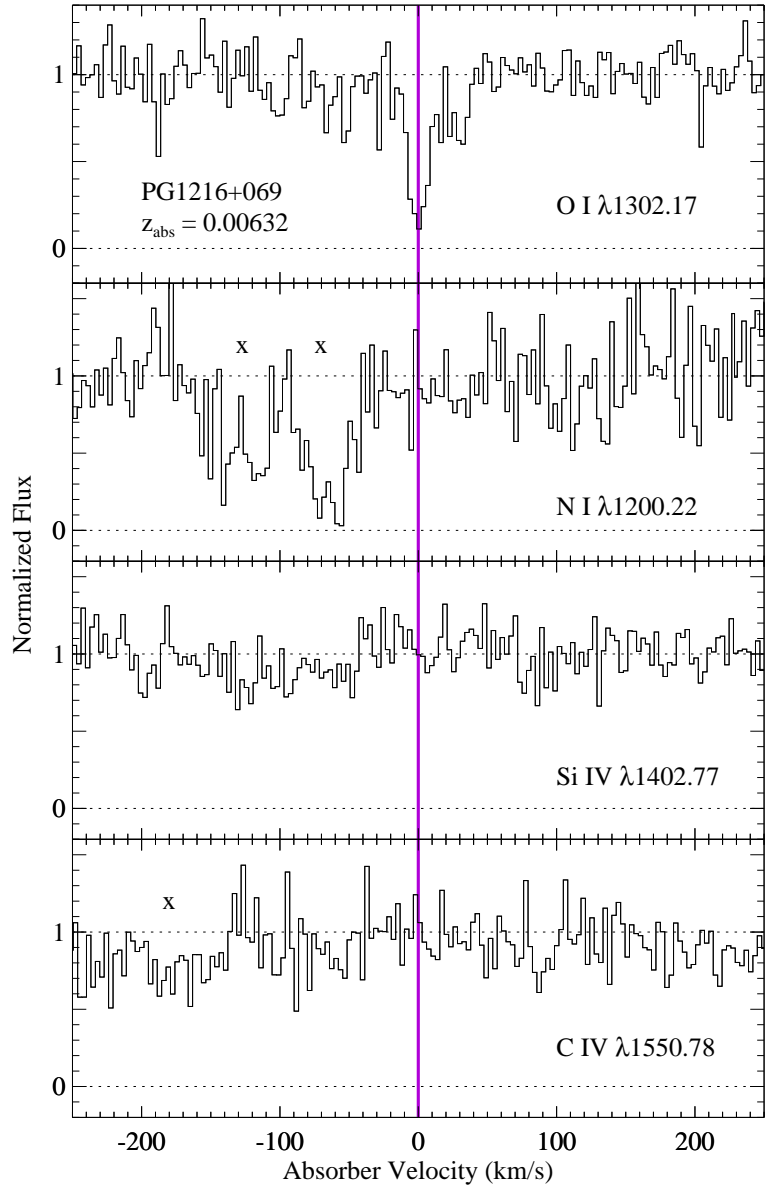}
\caption[]{Continuum-normalized regions of lines that are {\it not}
detected in the sub-damped \lya absorber at \zabs\ = 0.00632 (lower
three panels) with the \ion{O}{1} $\lambda$1302.17 profile replotted
in the top panel for purposes of comparison. As in
Figure~\ref{normprofs}, the data are plotted vs. velocity in the
absorber frame where $v = 0$ \kms\ at \zabs\ = 0.00632. Unrelated
absorption features from other redshifts are marked with an
'x'.\label{nocum}}
\end{figure}

Absorption lines of \ion{O}{1}, \ion{C}{2}, \ion{Si}{2}, and
\ion{Fe}{2} are detected at \zabs\ = 0.00632 in the STIS E140M
spectrum.  \ion{N}{1}, \ion{Si}{4}, and \ion{C}{4} are not apparent at
the 3$\sigma$ significance level. The strong \ion{Si}{3}
$\lambda$1206.50 line is redshifted into the core of the Galactic \lya
line and consequently cannot be reliably measured.\footnote{The region
of the spectrum that would encompass the \ion{Si}{3} feature was
excluded from the \ion{H}{1} fitting discussed in the previous
paragraph.} Selected absorption-line profiles of detected species are
shown in Figure~\ref{normprofs}; non-detections are plotted in
Figure~\ref{nocum}. We have measured the rest-frame equivalent widths
using the techniques of Sembach \& Savage (1992), which include
continuum placement uncertainty as well as a 2\% uncertainty in the
STIS flux zero level. These equivalent widths are listed in
Table~\ref{intprop} along with upper limits on undetected species of
interest.

We have used two techniques to estimate the column densities of metals
in the sub-DLA. First, we used standard Voigt-profile fitting with the
program of Fitzpatrick \& Spitzer (1997) and the line-spread functions
from the STIS Handbook (Kim Quijano et al. 2003). Inspection of the
absorption profiles (see Figure~\ref{normprofs}) reveals two
well-detected components at $v \approx$ 0 and 28 \kms\ in the absorber
velocity frame. The strongest lines show marginal evidence of
additional, barely detectable components at $-125 \lesssim v \lesssim
0$ \kms . These possible components are too poorly defined to justify
any attempts to fit them with Voigt profiles. Thus, we have elected to
only fit the two well-detected components; we discuss below how
inclusion of the additional weak features at $v \lesssim$ 0 \kms\
could affect our conclusions. Results from Voigt-profile fitting are
summarized in Table~\ref{voigtfit} including the velocity centroid,
$b-$value, and column density for the two primary components. 

For our second approach, we use the more generalized technique of
directly integrating the ``apparent column density'' profiles (see
Savage \& Sembach 1991; Jenkins 1996 and references therein). This
method does not incorporate any assumption that the velocity profile
is inherently Gaussian. Instead, each pixel is used to calculate the
apparent optical depth as a function of velocity, $\tau _{\rm a}(v) =
{\rm ln}[I_{\rm c}(v)/I(v)]$, where $I(v)$ is the observed intensity
and $I_{\rm c}(v)$ is the estimated continuum intensity at velocity
$v$. In turn, $\tau _{\rm a}(v)$ is used to calculate the apparent
column density, $N_{\rm a}(v) = (m_{\rm e}c/\pi
e^{2})(f\lambda)^{-1}\tau _{\rm a}(v)$.  If the profiles do not
contain narrow, saturated structures that are degraded by instrumental
smoothing, then \nav\ can be directly integrated to obtain a good
measurement of the total column density, $N_{\rm tot} = \int N_{\rm
a}(v) dv$. As in the equivalent width measurements, we used the
methods of Sembach \& Savage (1992) to incorporate the uncertainties
due to continuum placement and the flux zero point in the overall
\nav\ error bars.  The column densities from direct integration of
\nav\ profiles are given in Table~\ref{intprop}. We list both the {\it
total} column densities (integrated across both components) and the
integrated columns for the strongest component (at $v = 0$ \kms )
only.  We will make use of the measurements for the strongest
component only when we examine relative abundances in \S~\ref{relmet}.

It is important to note that several of the metal lines are strong and
narrow (see Figure~\ref{normprofs}), and we should be concerned about
whether unresolved saturation could cause the column densities to be
underestimated. Voigt profile fitting can adequately correct for
unresolved saturation to some extent by adjusting the line width, but
it has been shown that when lines are significantly affected by
unresolved saturation, profile fitting can produce erroneous results
(e.g., Shull et al. 2000; Sembach et al. 2001). The apparent column
density technique can also be used to check (and correct) for
unresolved saturation given two or more transitions of a particular
ion with significantly differing $f\lambda$ values. If the lines are
not affected by saturation, then the \nav\ profiles of weaker and
stronger transitions will agree. Conversely, if the profiles are
affected by saturated components, then weaker transitions will yield
significantly higher apparent column densities than corresponding
stronger transitions.

To check for saturation in the metal lines of the sub-DLA of this
paper, we made use of the fact that \ion{Si}{2} exhibits a generous
array of transitions within our spectral coverage. We detected the
\ion{Si}{2} $\lambda \lambda$ 1190.42, 1260.42, 1304.37, and 1526.71
transitions (the 1193.29 \AA\ line could not be used due to strong
blending with Galactic absorption); the weakest and strongest lines in
this set differ by a factor of 12 in $f\lambda$ (see
Table~\ref{intprop}). When we compared the apparent optical depths of
the different \ion{Si}{2} lines, it was clear that they did not scale
in proportion to the values of $f\lambda$, indicating the presence of
unresolved saturation. A method to correct \nav\ profiles for
unresolved saturation has been outlined by Jenkins (1996).  The
discussion in Jenkins (1996) applies to the analysis of doublets, but
the arguments can be generalized to include simultaneously many lines
of differing strength.  We have applied this method to the sub-DLA
\ion{Si}{2} lines in order to correct for the unresolved
saturation. At each velocity, we solved for the minimum values of the
$\chi^2$ in all of the recorded intensities by allowing two parameters
to vary: the true column density per unit velocity $N(v)$ and a
saturation parameter, one that is directly analogous to the $b-$value
in a classical application of a curve of growth to measurements of
equivalent widths.  Much of the detail in the final outcome is
governed by the strongest line, since it has the greatest sensitivity
to small changes in $N(v)$.  Nevertheless, the weaker lines are able
to influence the shape of the profile where the stronger lines are
affected by saturation, creating an outcome that corrects for the
distortion caused by unresolved, saturated structures. This method
also automatically gives the greatest weight to the most useful
transition for each velocity element. That is, moderate optical depths
prevail over the weak ones buried in the noise or the very strong ones
that are more subject to systematic errors.

\begin{figure}
\epsscale{1.0}
\plotone{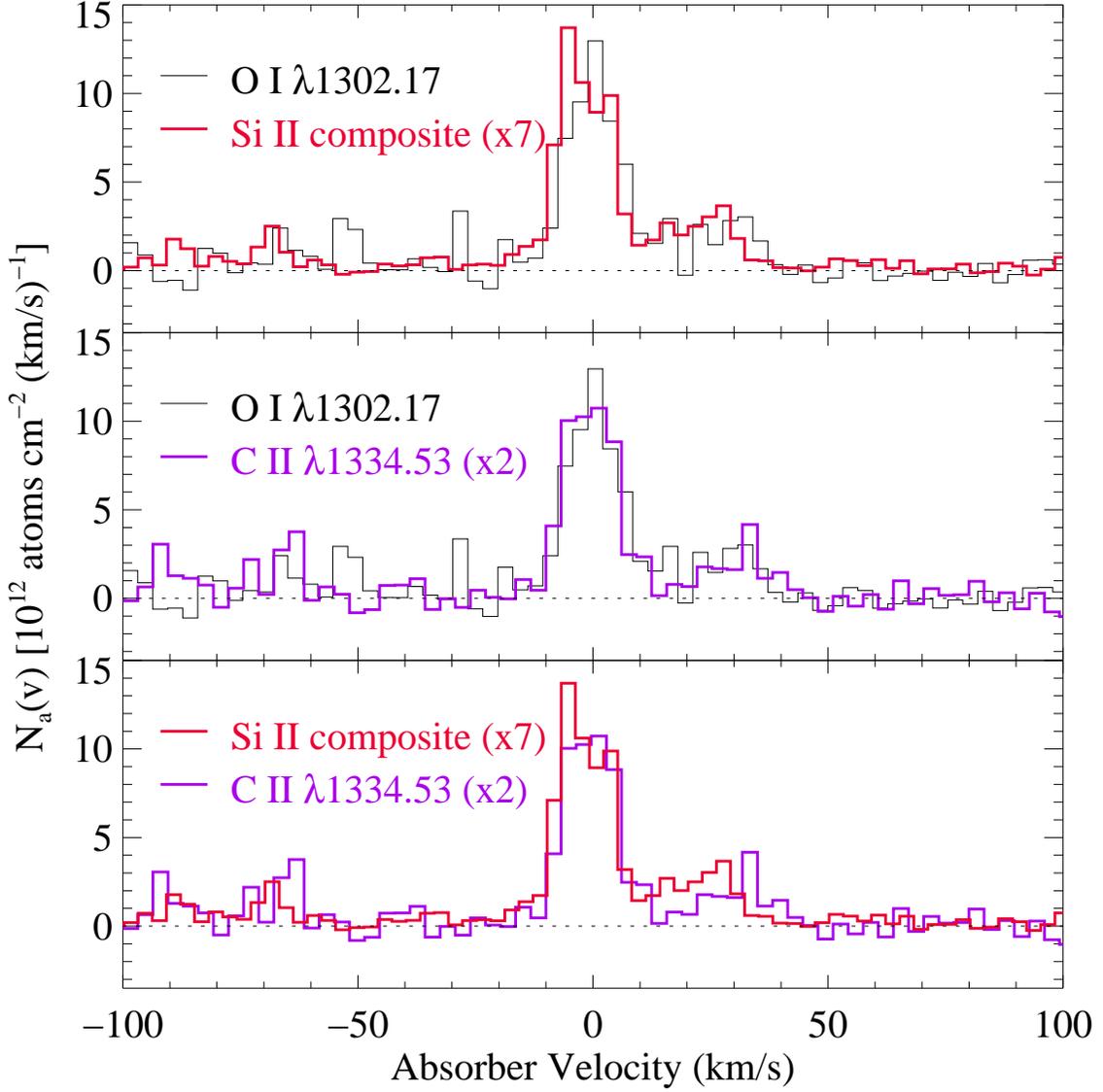}
\caption[]{Comparison of apparent column density profiles, \nav , of
various species detected in the sub-DLA at \zabs\ = 0.00632: (top)
\ion{O}{1} $\lambda$1302.17 (thin line) vs. a composite \ion{Si}{2}
profile (thick line) constructed from the \ion{Si}{2} $\lambda
\lambda$1190.42, 1260.42, 1304.37, and 1526.71 transitions (see text),
(middle) \ion{O}{1} $\lambda$1302.17 (thin line) vs. \ion{C}{2}
$\lambda$1334.53 (thick line), (bottom) \ion{Si}{2} composite (thin
line) vs. \ion{C}{2} $\lambda$1334.53 (thick line).  In all panels,
the \ion{Si}{2} and \ion{C}{2} profiles have been scaled up by factors
of 7 and 2, respectively, and the velocities have been corrected as
discussed in the appendix.\label{navplot}}
\end{figure}

The saturation-corrected, composite \ion{Si}{2} profile is compared to
the \ion{O}{1} $\lambda$1302.17 and \ion{C}{2} $\lambda$1334.53
profiles in Figure~\ref{navplot}.  The correction implied by the
\ion{Si}{2} \nav\ profiles turned out to be relatively small, and the
total \ion{Si}{2} column derived from integration of the
saturation-corrected composite \ion{Si}{2} profile is log
$N$(\ion{Si}{2}) = 13.52. This saturation-corrected total \ion{Si}{2}
column is in excellent agreement with the results from profile fitting
(summing the two components, see Table~\ref{voigtfit}), which suggests
that the profile fitting properly accounts for the saturation in the
case of \ion{Si}{2} since it provides the best match to all
transitions simultaneously. What about the \ion{O}{1} and \ion{C}{2}
profiles, for which we do not have the benefit of multiple transitions
for saturation checking? It is possible that the \ion{O}{1} and
especially the \ion{C}{2} profiles are affected by unresolved
saturation, but the effects of saturation are likely to be modest. We
note that the shapes of the \ion{O}{1} and \ion{C}{2} \nav\ profiles
are in good agreement with the saturation-corrected \ion{Si}{2}
profile in the main (strongest) component.\footnote{Some small
discrepancies are evident in the weaker component at $v \approx$ 25
\kms\ in Figure~\ref{navplot}. While these differences could be due to
inadequate correction for the wavelength calibration errors noted in
the appendix or could be real differences, given the weakness of this
component, the disagreement could be due to noise or the possible
presence of a warm pixel at the red edge of the \ion{Si}{2} $\lambda
1260.42$ profile.}  Moreover, the \ion{O}{1} and \ion{C}{2} $b-$values
derived from profile fitting are in excellent agreement with the
\ion{Si}{2} $b-$values (see Table~\ref{voigtfit}); the profile-fitting
results likely correct for the saturation adequately well.  Good
abundances based on \ion{O}{1} are particularly valuable, but we will
show below that we reach the same essential conclusions for the
abundances of the sub-DLA if we build our analysis mainly on the
saturation-corrected \ion{Si}{2} measurements, which are robust
against this source of systematic error. We note that the \ion{Fe}{2}
column density from profile fitting is 0.2 dex higher than
$N$(\ion{Fe}{2}) obtained from \nav\ integration. However, the
\ion{Fe}{2} measurements are the noisiest in the set of detected
lines, and in fact the \ion{Fe}{2} columns from the two techniques
agree within their 1$\sigma$ uncertainties.

\section{Abundances and Dust Content\label{abund}}

We now turn to the abundances implied by the column densities
contained in Tables~\ref{h1col}-\ref{voigtfit}. We begin with the
absolute metallicity of the absorber (\S~\ref{absmet}), and then we
examine the patterns of relative abundances from element to element
and explore their implications regarding the dust content and
nucleosynthetic history of the sub-DLA (\S~\ref{relmet}). Throughout
this paper, we use the solar oxygen and carbon abundances from Allende
Prieto et al. (2001,2002), and we take the N, Si, and Fe solar
abundances from Holweger (2001).

\subsection{Absolute Abundance\label{absmet}}

Abundance measurements in DLAs and sub-DLAs can suffer from uncertain
corrections for ionization (e.g., Howk \& Sembach 1999; Prochaska et
al. 2002a) and depletion by dust (Prochaska 2003). It is well known
that these problems can be largely avoided by using \ion{O}{1} to
determine the absolute metallicity of an absorber because (1) the
ionization potentials of \ion{O}{1} and \ion{H}{1} are very similar,
and \ion{O}{1} is strongly tied to \ion{H}{1} by resonant charge
exchange (Field \& Steigman 1971); consequently, the ionization
correction is so small that it can largely be neglected for
\ion{O}{1}, and (2) evidence from the local region of our Galaxy
indicates that oxygen is only lightly depleted by dust (Meyer et
al. 1998; Moos et al. 2002; Andr\'{e} et al. 2003). Our measurement of
$N$(\ion{O}{1}) is therefore particularly advantageous for
constraining the metallicity of the \zabs\ = 0.00632 absorber. Using
the summed \ion{O}{1} column from Voigt-profile fitting for the two
main components (see Table~\ref{voigtfit}) and $N$(\ion{H}{1}) from
the \lya fit, we obtain
\begin{equation}
[{\rm O/H}] = -1.60^{+0.09}_{-0.11}\label{absabun}
\end{equation}
where we have used the usual logarithmic notation.\footnote{[X/Y] =
log (X/Y) - log (X/Y)$_{\odot}$.}

There are some additional sources of uncertainty that are not
incorporated into the error bar in eqn.~\ref{absabun}. We noted in
\S~\ref{abmeas} that additional weak components may be present at $v
\leq 0$ \kms .  These components are unlikely to be saturated by
virtue of their weakness.  Integrating these extra features in the
\ion{O}{1} $\lambda$1302.17 and \ion{Si}{2} $\lambda$1260.42 profiles
out to $v = -125$\kms , we find that the \ion{O}{1} column increases
by 0.11 dex and $N$(\ion{Si}{2}) increases by 0.09 dex.  Therefore
these extra components have only a small effect; inclusion of these
components increases the metallicity by $\sim$0.1 dex.

Another possible source of uncertainty in the oxygen abundance is the
possibility that saturation has led to a substantial underestimate of
$N$(\ion{O}{1}).  We have argued that saturation is unlikely to be a
major problem (\S~\ref{abmeas}), but we can alternatively constrain
the metallicity using the saturation-corrected \ion{Si}{2}
measurements. However, \ion{Si}{2} can require significant ionization
corrections (see, e.g., \S 5 in Tripp et al. 2003). To evaluate the
ionization correction required for \ion{Si}{2}, we have employed
CLOUDY photoionization models (v94, Ferland et al. 1998) as described
in Tripp et al. (2003). In these models, the gas is photoionized by
the UV background from quasars at $z \approx 0$. The intensity of the
UV background at 1 Rydberg is taken to be $J_{\nu} = 1 \times
10^{-23}$ ergs s$^{-1}$ cm$^{-2}$ Hz$^{-1}$ sr$^{-1}$ based on
observational constraints (Weymann et al. 2001; Dav\'{e} \& Tripp
2001, and references therein) with the radiation field shape computed
by Haardt \& Madau (1996). The absorber is approximated as a
constant-density, plane-parallel slab, and the thickness of the slab
is adjusted to reproduce the observed \ion{H}{1} column. 

For our modeling of the ionization, we focused our attention on the
main, strong component at $v = 0$; \ion{Fe}{2} is not detected in the
weaker component, and we obtain tighter contraints on \ion{N}{1} by
integrating over the stronger component only. Assuming the
metallicities are similar in the strong and weak components, we
reduced $N$(\ion{H}{1}) in the strong component-only photoionization
models since some of the \ion{H}{1} is surely located in the weaker
component.  Based on the strong/weak component metal column density
ratios, we estimate that log $N$(\ion{H}{1}) = 19.23 in the strong
component only, and we shall see that this provides a good fit to the
detected metals in this component. Figure~\ref{photmodel} shows the
column densities of \ion{O}{1}, \ion{N}{1}, \ion{C}{2}, \ion{Si}{2},
\ion{Fe}{2}, \ion{C}{4}, and \ion{Si}{4} predicted by the
photoionization model as a function of the ionization parameter $U$
($\equiv$ ionizing photon density/total H number density $=
n_{\gamma}/n_{\rm H}$) compared to the observed column densities and
upper limits (for the $v = 0$ \kms\ component only). The relative
metal abundances (e.g., O/Si) are fixed to the solar values according
to the references at the beginning of this section and assume no
depletion by dust. The model predictions are shown by solid curves
with small symbols, and the observed column densities are indicated
with large symbols with $1\sigma$ error bars ($3\sigma$ upper limits
are indicated with arrows). With the exceptions of \ion{N}{1} and
\ion{Fe}{2}, we see that the model is in good agreement with all
detections and upper limits at log $U \approx -3.8$. The model agrees
with the \ion{Fe}{2} column from direct integration at this $U$ as
well, but $N$(\ion{Fe}{2}) from profile-fitting is 0.2 dex higher than
predicted. The \ion{Fe}{2} line is narrow and marginally detected, and
consequently this discrepancy could easily be due to noise.  We will
discuss the origin of the discrepancy between the model and the
observations in the case of \ion{N}{1} below. For the assumed UV
background shape and $J_{\nu}$ value, log $U = -3.8$ corresponds to
log $n_{\rm H} = -2.6$ (however, $n_{\rm H}$ is not well-constrained
because it depends on the assumed intensity of the radiation field,
which is only loosely constrained; see \S~\ref{physcon}).  The absence
of \ion{C}{4} and \ion{Si}{4} provide upper limits on the ionization
parameter, and these upper limits translate to lower limits on the Si
abundance. As $U$ increases, $N$(\ion{Si}{2}) predicted by the model
increases (see Figure~\ref{photmodel}), so the metallicity must be
decreased accordingly to match the observed \ion{Si}{2} column.  At
the left edge of Figure~\ref{photmodel}, the curves each approach an
asymptote for a simple reason: as $U$ decreases, the ionization
corrections become negligible, and for ions that are dominant in
neutral gas, the abundance can then be calculated directly from the
ion/\ion{H}{1} ratio, e.g., [Si/H] = [\ion{Si}{2}/\ion{H}{1}].  This
condition places the upper limit on [Si/H]. With the upper limit on
$U$ required by the \ion{Si}{4} constraint, we find
\begin{equation}
-1.77 \leq [{\rm Si/H}] \leq -1.35.\label{siabun}
\end{equation}
The range in eqn.~\ref{siabun} only reflects the uncertainty in the
ionization correction; the additional uncertainty due to the error
bars in the column densities and solar Si abundance amounts to
$\pm0.08$ dex. The metallicity implied by the \ion{Si}{2} measurements
is therefore similar to the absolute abundance derived from
\ion{O}{1}.

The low absolute abundances derived from the oxygen and silicon
measurements are notable given the redshift of the absorber and its
location in the outskirts of the Virgo cluster region and NGC4261
group (see Figure~\ref{xraymap}). As we show in Figure~\ref{metcomp},
this sub-DLA has an absolute metallicity that is similar to comparable
sub-DLAs and DLAs at high redshifts. Figure~\ref{metcomp} compares the
oxygen abundance of the PG1216+069 sub-DLA (large filled circle) to
the mean metallicities derived from 125 DLAs at 0.5 $<$ \zabs\ $<$ 5
in six redshift bins (open squares; Prochaska et al. 2003) and
individual measurements of oxygen abundances in sub-DLAs at \zabs\ $>$
1.7 (small filled circles; Dessauges-Zavadsky et
al. 2003).\footnote{For the reasons outlined at the beginning of
\S4.1, we have restricted our comparison to [O/H] measurements for
high$-z$ sub-DLAs, such as those from Dessauges-Zavadsky et
al. (2003). In the higher-$N$(\ion{H}{1}) DLAs, the oxygen lines are
often badly saturated and/or difficult to measure due to blending with
\lya forest lines. Consequently, the DLA abundances are based on Si,
S, O, or Zn measurements; see Prochaska et al. (2003) for further
information.}  Evidently the Virgo sub-DLA metallicity is equal to or
less than the metallicity of many of the high-redshift absorbers. This
suggests that this low-z absorber is a relatively primitive gas cloud.

On the other hand, the Virgo sub-DLA metallicity {\it is}
significantly higher than the typical metallicities derived in studies
of low$-\delta$ Ly$\alpha$ forest clouds at \zabs $\gtrsim$ 2 (e.g.,
Schaye et al. 2003; Simcoe et al. 2004).  This suggests that the
PG1216+069 sub-DLA has somehow received at least some subsequent
injection of metals above and beyond the first wave of enrichment that
produced the widespread distribution of carbon and oxygen seen in the
high-z forest.  There are few constraints available on the metallicity
of Ly$\alpha$ forest clouds at the present epoch. However, some
low$-z$ absorbers with \ion{H}{1} column densities in the range
expected for Ly$\alpha$ forest clouds have metallicities that are
considerably higher (e.g., Tripp \& Savage 2000; Savage et al. 2002)
than that of the Virgo sub-DLA.  Therefore while the Virgo absorber is
apparently not as pristine as the high$-z$ Ly$\alpha$ forest, it
nevertheless is relatively metal-poor compared to analogous objects in
the nearby universe.

\subsection{Relative Abundances and Dust\label{relmet}}

Beyond the overall metallicity, we have two motivations for
scrutinizing the detailed pattern of element-to-element abundances in
the sub-DLA. First, relative abundances can provide valuable
constraints on the nucleosynthetic history of the object, as has been
widely discussed in the stellar abundance and QSO absorption-line
literature (e.g., Lauroesch et al. 1996; McWilliam 1997). Second,
relative abundances can reveal the presence of dust through the
patterns of differential depletion (e.g., Savage \& Sembach 1996;
Jenkins 2003). Unfortunately, for many elements the nucleosynthesis
signatures and dust depletion patterns can be degenerate (Prochaska
2003). For this reason, nitrogen abundances are particularly valuable.
Nitrogen nucleosynthesis depends on the initial metallicity of the
gas, and consequently in ``chemically young'' gas, nitrogen is
expected (and observed) to be significantly underabundant (e.g. Vila
Costas \& Edmunds 1993; Pettini et al. 1995; Henry et
al. 2000). Nitrogen is only lightly depleted by dust (Meyer et
al. 1997), so N/O provides a probe of the nucleosynthetic history that
is not highly confused by dust depletion. Good iron abundances are
also particularly useful because Fe is highly prone to depletion by
dust, even in low-density, halo gas clouds (e.g., Savage \& Sembach
1996). Consequently, iron can reveal the presence of dust when many
other elements show no depletion (see further discussion in Jenkins
2003).

\begin{figure}
\epsscale{1.0}
\plotone{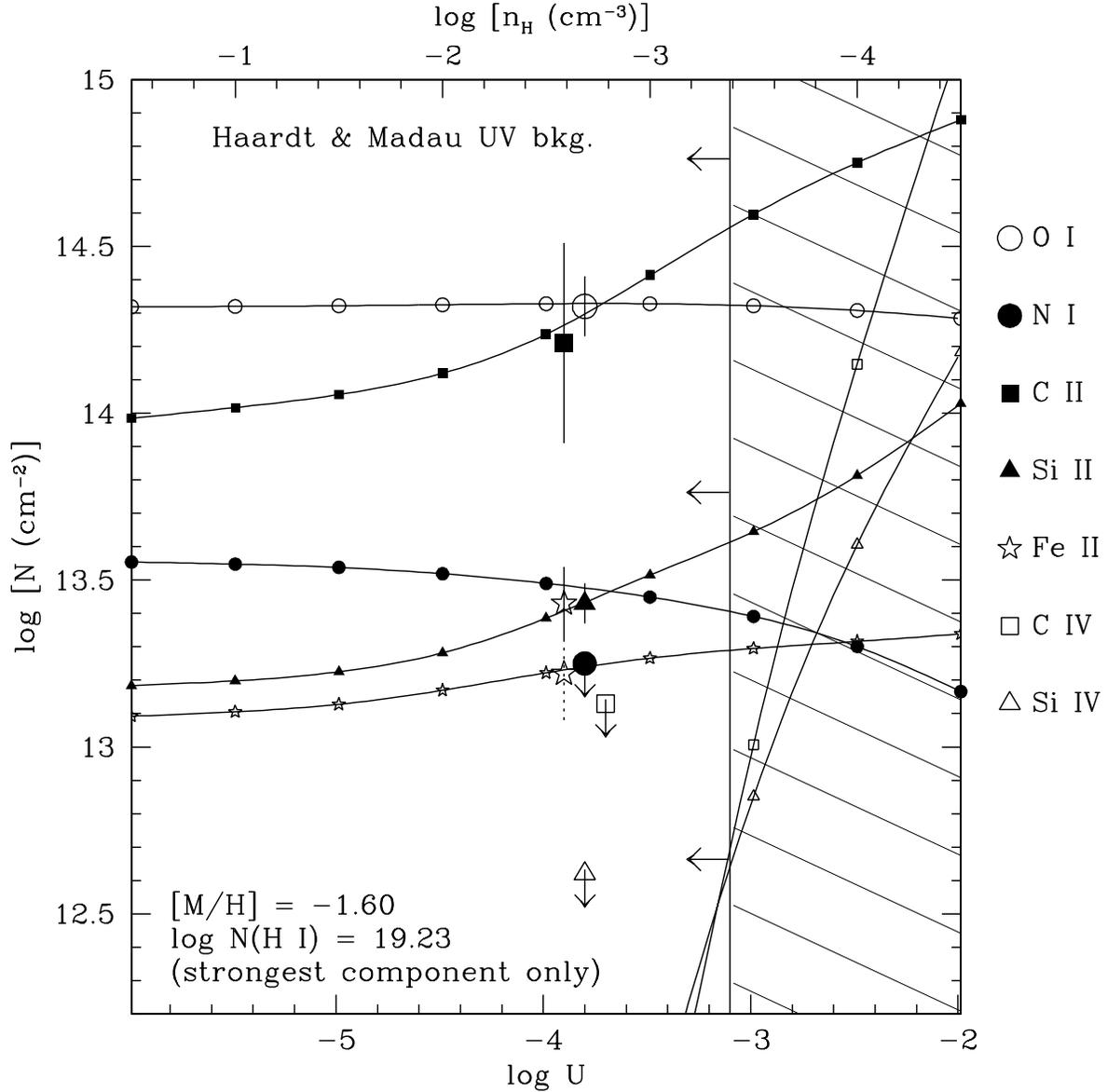}
\caption[]{\footnotesize Comparison of column densities predicted by a
photoionization model (solid curves with small symbols) to the
observed columns in the stronger component of the sub-DLA at \zabs\ =
0.00632 (large symbols). The species corresponding to each symbol are
shown in the key at right. The model column densities are plotted as a
function of the ionization parameter $U$ (lower axis, $U \equiv
n_{\gamma}/n_{\rm H}$) and the hydrogen number density (upper
axis). The model assumes the Haardt \& Madau (1996) UV background at
$z = 0$ with $J_{\nu} = 1 \times 10^{-23}$ ergs cm$^{-2}$ s$^{-1}$
Hz$^{-1}$ sr$^{-1}$. The observed column densities are plotted near
the best-fitting value of log $U \approx -3.8$, and points with arrows
are 3$\sigma$ upper limits from direct integration. For most of the
well-detected species, we use the columns from component fitting in
this plot. For iron, which is based on a single weak and narrow line,
we show both the component-fitting measurement (upper point with solid
error bars) and the column density from direct integration (lower
point with dotted error bars). The model \ion{Si}{4} and \ion{C}{4}
columns exceed the observed 3$\sigma$ upper limits in the hatched
portion of this figure; the \ion{Si}{4} constraint requires log $U
\leq -3.1$.\label{photmodel}}
\end{figure}

\begin{figure}
\epsscale{1.0}
\plotone{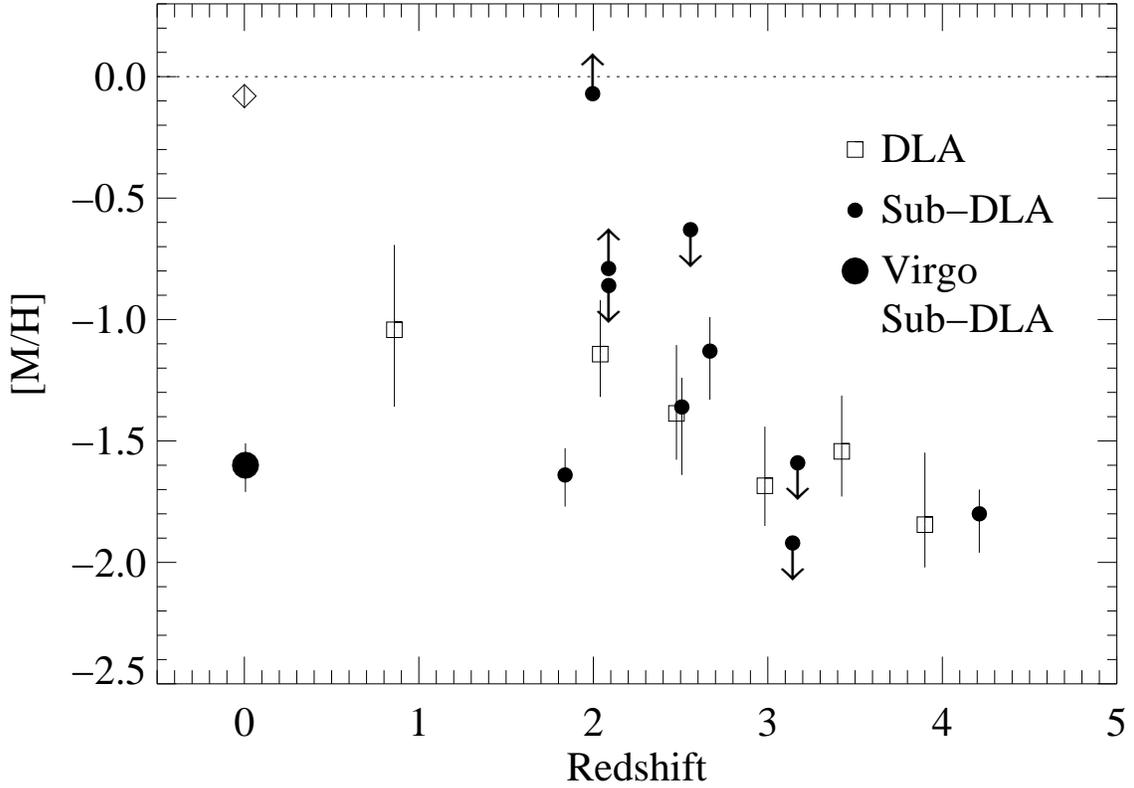}
\caption[]{Logarithmic metallicity in the Virgo sub-DLA at \zabs\ =
  0.00632 (large filled circle) compared to oxygen abundances in
  sub-DLAs at high redshifts (small filled circles) from the sample of
  Dessauges-Zavadsky et al. (2003) and the binned, unweighted mean
  metallicities from the sample of 125 DLAs (open squares) compiled by
  Prochaska et al. (2003). The DLA bins are shown with 95\% confidence
  limits, and the bins at $z > 1.5$ contain equal numbers of DLA
  absorbers; the lowest-redshift bin contains fewer systems due to the
  paucity of low$-z$ DLAs that have been found and measured. The
  diamond shows the mean oxygen abundance in the Milky Way ISM in the
  general vicinity of the Sun (Andre et al. 2003).\label{metcomp}}
\end{figure}

We begin our examination of the relative abundances with iron.  We
have only detected a single \ion{Fe}{2} line, and moreover the line is
weak and narrow.  The \ion{Fe}{2} column densities from direct
integration and profile fitting agree to within their 1$\sigma$
uncertainties, but the best values from the two techniques differ by
0.2 dex. If we neglect ionization corrections, we find [Fe/O] =
0.15$^{+0.15}_{-0.25}$ or [Fe/O] = 0.35$^{+0.15}_{-0.23}$ using the
columns from direct integration and profile fitting,
respectively. Dust depletion would lead to iron underabundance, but in
this case the opposite trend is seen: Fe is marginally overabundant
with respect to oxygen.  Therefore, the iron measurements indicate
that {\it this absorption system contains very little dust}. In
principle, \ion{Fe}{2} can also require substantial ionization
corrections. However, it can be seen from Figure~\ref{photmodel} that
with the upper limit on $U$ imposed by the high ions, the \ion{Fe}{2}
ionization correction is small. Indeed, the ionization model in
Figure~\ref{photmodel} adequately reproduces the column densities of
all detections and limits (except the \ion{N}{1} limit), without
requiring any depletion, at log $U \approx -3.8$.

\begin{figure}
\epsscale{1.0}
\plotone{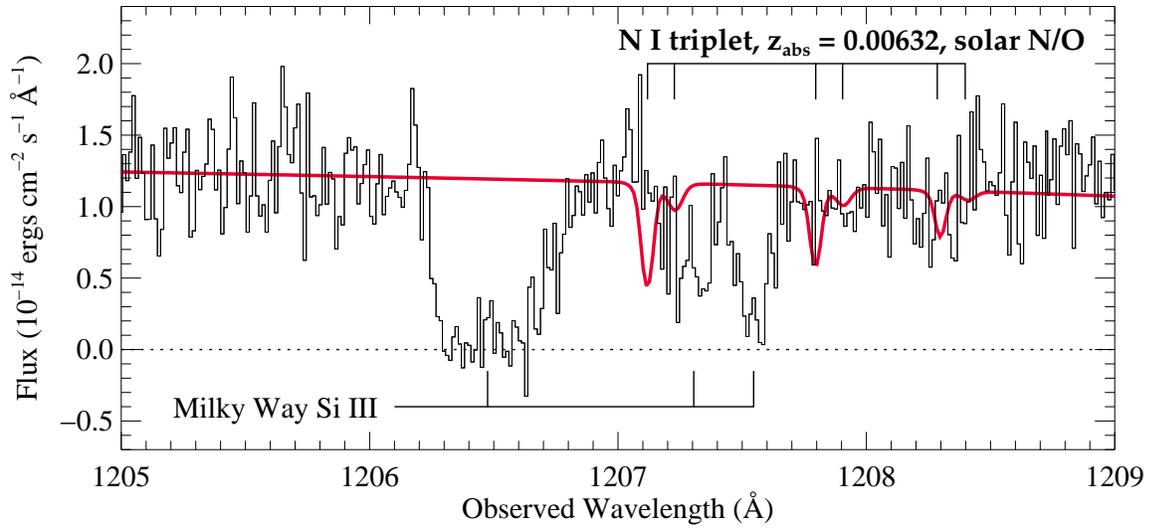}
\caption[]{Comparison of the predicted absorption profiles of the
\ion{N}{1} $\lambda \lambda$1199.55, 1200.22, 1200.71 triplet at
\zabs\ = 0.00632, {\it assuming the relative N/O abundance is solar},
to the observed spectrum of PG1216+069. The \ion{N}{1} triplet is
redshifted into a noisy region of the spectrum, and moreover, it is
blended with \ion{Si}{3} $\lambda$1206.50 absorption from
high-velocity Milky Way gas. Nevertheless, the predicted \ion{N}{1}
profiles are clearly too strong; nitrogen is underabundant in this
absorber.\label{nitrounder}}
\end{figure}

Using the 3$\sigma$ upper limit on $N$(\ion{N}{1}) from
Table~\ref{intprop} and neglecting ionization corrections, we find
[N/O] $\leq -0.28$ in the strongest component of the PG1216+069
sub-DLA. While this \ion{N}{1} underabundance could be due to
ionization effects (e.g., Sofia \& Jenkins 1998), the ionization model
in Figure~\ref{photmodel} shows that this is probably not due to
ionization in this case if the gas is photoionized by the background
radiation from QSOs. The model in Figure~\ref{photmodel}, which
assumes that the relative abundance of nitrogen to oxygen is solar,
predicts a neutral nitrogen column that is $\sim$0.25 dex higher than
the 3$\sigma$ upper limit on $N$(\ion{N}{1}) at the best value of $U$.
Increasing $U$ decreases this discrepancy, but at the upper limit on
$U$ set by the absence of \ion{Si}{4}, the model still exceeds the
3$\sigma$ upper limit (see Figure~\ref{photmodel}). We conclude that
the paucity of \ion{N}{1} reflects a true nitrogen underabundance.
Our upper limit on $N$(\ion{N}{1}) in Table~\ref{intprop} only assumes
that the linear curve of growth is applicable; no other assumptions
about the $b-$value were made. Figure~\ref{nitrounder} shows the
predicted \ion{N}{1} profiles for both of the main components, if we
assume that the N/O ratio is solar and that the \ion{N}{1} components
have the same $b-$value as the detected \ion{O}{1}
components. Although the region of the strongest \ion{N}{1} transition
is close to a few warm pixels, it is nevertheless evident from
Figure~\ref{nitrounder} that a solar N/O abundance does not agree with
the data.

But what if the gas is also photoionized by {\it embedded} stars that
have a much softer spectrum than the background flux employed in
Figure~\ref{photmodel}? It is conceivable that the outer envelope of
\ion{H}{1} could provide enough shielding to prevent penetration of
the absorber by photons with appropriate energy\footnote{High-energy
photons with energies substantially greater than the ionization
potentials of \ion{Si}{3} and \ion{C}{3} {\it can} penetrate a
sub-DLA/DLA, but as can be seen from the model in
Figure~\ref{photmodel}, which includes such photons, they are
inefficient at production of \ion{Si}{4} and \ion{C}{4}.} to produce
significant amounts of \ion{Si}{4} and \ion{C}{4}, thereby satisfying
the observational upper limits on these species, but lower-energy
photons from internal stars could still ionize the nitrogen leading to
a deficit of \ion{N}{1}. In high-redshift DLAs, a striking
correspondence between the absorption-profile structure of \ion{Al}{3}
and lower-ionization stages\footnote{The ionization potential of
\ion{Al}{2} is 18.8 eV, so the correlation of \ion{Al}{3} and
lower-ionization stages indicates that ionized gas is present and
associated with the putatively neutral gas. Unlike the Virgo sub-DLA
observed toward PG1216+069, the high$-z$ DLA and sub-DLA systems often
show strong high-ion absorption from species such as \ion{Si}{4} and
\ion{C}{4}, but usually the \ion{Si}{4}/\ion{C}{4} component structure
is significantly different from that of the low ions and \ion{Al}{3}
(see, e.g., Lu et al. 1996; Dessauges-Zavadsky et al. 2003).} has been
noted; this could indicate that soft stellar spectra play a role in
the ionization of the high$-z$ DLAs (e.g., Howk \& Sembach
1999). Also, in the Milky Way ISM within the ``Local Bubble'' in the
immediate vicinity of the Sun, a deficit of \ion{N}{1} has been
observed (Jenkins et al. 2000; Lehner et al. 2003); in this context,
the low \ion{N}{1} columns are quite probably due to ionization
effects since other observations of sight lines extending beyond the
Local Bubble boundaries suggest that the interstellar nitrogen
abundance is rather uniform in the solar neighborhood, and moreover
nitrogen in this region is not underabundant (e.g., Meyer et
al. 1997).

\begin{figure}
\epsscale{1.1}
\plotone{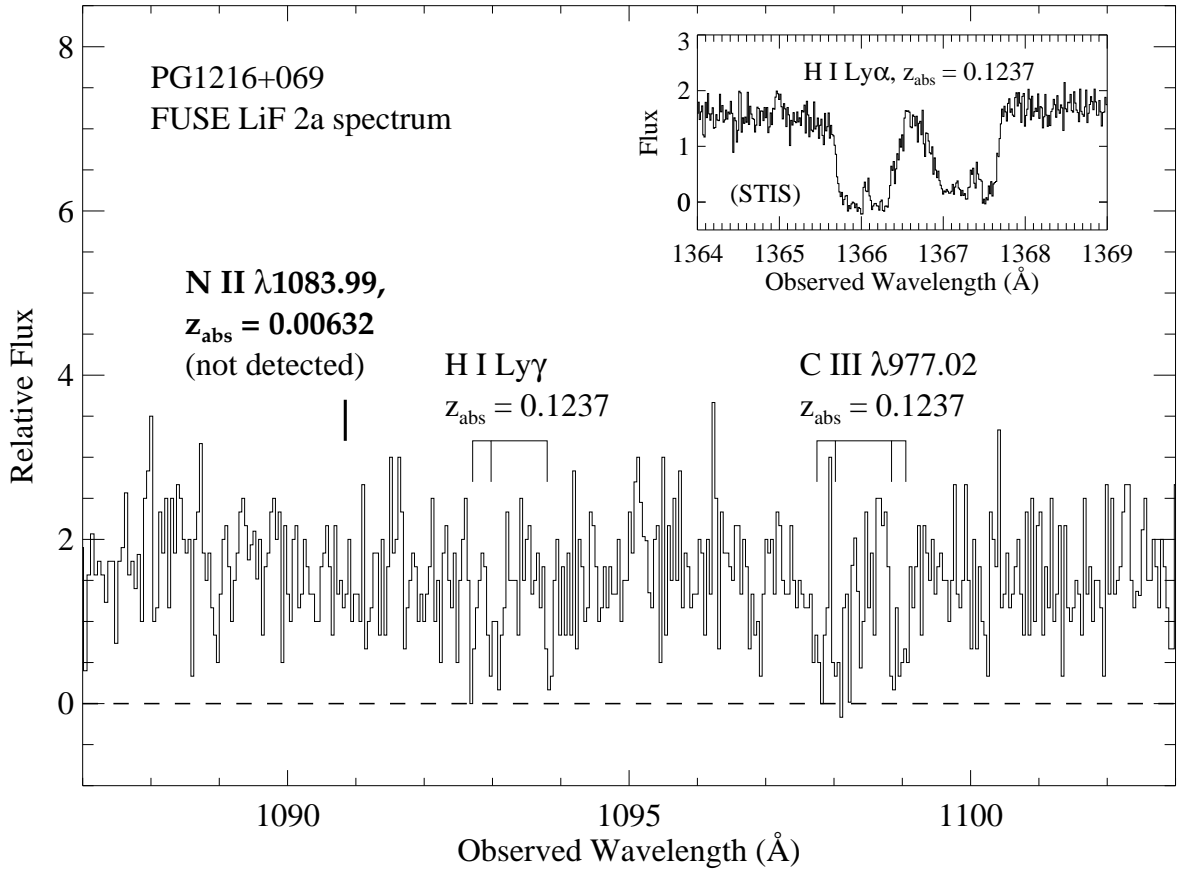}
\caption[]{Portion of the {\it FUSE} spectrum of PG1216+069 covering the
expected wavelength of the \ion{N}{2} $\lambda$1083.99 transition at
\zabs\ = 0.00632. The spectrum is plotted vs. observed wavelength, and
the redshifted wavelength of the \ion{N}{2} line is indicated; no
statistically significant feature is apparent at the expected
position. The {\it FUSE} spectrum is relatively noisy, but
nevertheless multiple Ly$\gamma$ and \ion{C}{3} $\lambda$977.02 lines
from the strong, multicomponent absorber at \zabs\ = 0.1237 are
clearly detected; these lines are also marked. To corroborate the
identification of these Ly$\gamma$ and \ion{C}{3} lines, the inset
shows the multicomponent Ly$\alpha$ profile at \zabs\ = 0.1237 from
the STIS spectrum; several of the components in the \lya profile are
also evident in the Ly$\gamma$ and \ion{C}{3} transitions. This figure
shows only the LiF2a segment.  The only other segment of the {\it
FUSE} spectrum that covers this region (the SiC2b segment) is too
noisy to provide any useful information.\label{fusen2spec}}
\end{figure}

Considering these observations of high$-z$ DLAs and the Local Bubble
ISM, we have multiple motivations to investigate whether ionization by
light from embedded stars could be responsible for the paucity of
\ion{N}{1} in the PG1216+069 sub-DLA. The most straightforward way to
test this hypothesis is to search for the \ion{N}{2} $\lambda$1083.99
absorption line at the sub-DLA redshift; if the lack of \ion{N}{1} is
due to ionization from soft sources, then \ion{N}{2} should be
correspondingly strong.  Figure~\ref{fusen2spec} shows the portion of
the {\it FUSE} spectrum covering the \ion{N}{2} $\lambda$1083.99
transition at \zabs\ = 0.00632. No significant \ion{N}{2} absorption
is apparent at the expected wavelength. The signal-to-noise ratio of
the {\it FUSE} spectrum is not overwhelming, but the clear detection
of absorption lines from other redshifts (see also
Figure~\ref{fuseo6spec}) demonstrates that the spectrum is adequate to
reveal \ion{N}{2} at the expected strength. Due to the lower spectral
resolution of {\it FUSE}, the two main components of the sub-DLA are
blended, so we have integrated over both components to place limits on
\ion{N}{2}. We find that $W_{\rm r} \leq$ 140.4 m\AA\ at the 3$\sigma$
level. Assuming the linear curve-of-growth applies, this corresponds
to log $N$(\ion{N}{2}) $\leq$ 14.07.

What does the \ion{N}{2} limit imply? We have computed several CLOUDY
photoionization models assuming various soft ionizing radiation fields
including stellar ionizing radiation approximated by a Kurucz model
atmosphere with either $T_{\rm eff}$ = 30,000 or 50,000 K as well as
an estimate of the interstellar radiation field in the disk of the
Milky Way (Black 1987). Of these three models, only the Kurucz stellar
atmophere spectrum with $T_{\rm eff}$ = 50,000 K satisfies the
observed upper limit on \ion{N}{1}/\ion{O}{1} in the Virgo sub-DLA
(the other two models cannot produce a low enough
\ion{N}{1}/\ion{O}{1} ratio). However, when the \ion{N}{1}/\ion{O}{1}
limit is satisfied, the 50,000 K stellar spectrum produces an
\ion{N}{2} column density that exceeds our 3$\sigma$ upper
limit. Based on these models and the \ion{N}{2} limit, we conclude
that internal ionization by soft sources is also an unlikely
explanation of the absence of nitrogen absorption in the PG1216+069
sub-DLA. 

\begin{figure}
\epsscale{1.0}
\plotone{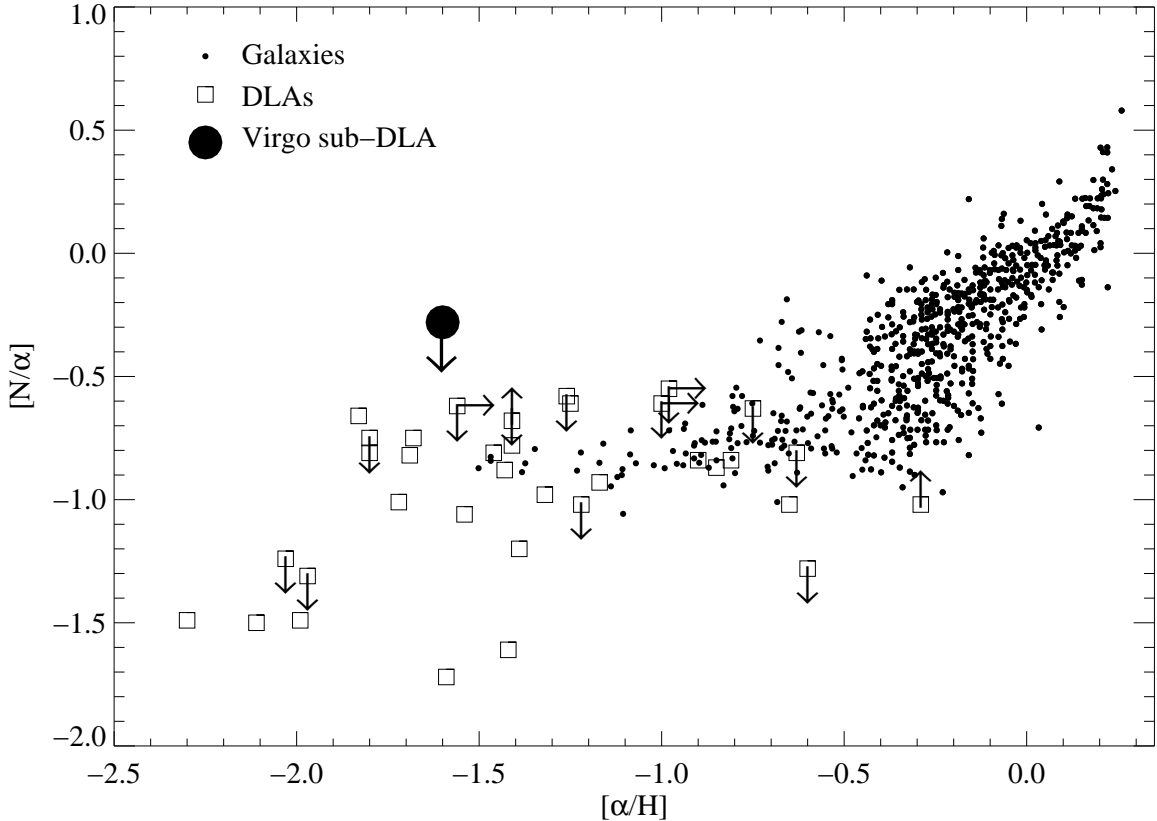}
\caption[]{Nitrogen underabundance and absolute oxygen abundances
measured in the PG1216+069 sub-DLA (large filled circle) compared to
nitrogen abundance patterns observed in low-redshift galaxies (small
filled circles) and high-z damped \lya absorbers (open squares). The
x-axis indicates the absolute $\alpha -$element abundance, and the
y-axis shows the relative nitrogen abundance with respect to an
$\alpha -$element.  As discussed in \S ~\ref{absmet}, oxygen provides
the most reliable measurement of [$\alpha$/H] and [N/$\alpha$], so we
prefer to use oxygen measurements when they are available.  However,
in some cases the oxygen lines are excessively saturated or not
observed, in which case sulfur or silicon were used to estimate the
$\alpha$ abundance. The galaxy \ion{H}{2} region data were compiled by
Pilyugin et al. (2003) and converted to abundances using the
``P-method''.  The high-z DLA measurements are from Centuri\'{o}n et
al. (2003) with the exceptions and additions detailed by Jenkins et
al. (2004).\label{undern}}
\end{figure}

The finding that nitrogen is underabundant also indicates that this
sub-DLA is a primitive and chemically young system. In
Figure~\ref{undern} we compare the Virgo sub-DLA nitrogen and oxygen
abundances to measurements from a large sample of \ion{H}{2} regions
in nearby galaxies (small filled circles) and high-z DLAs (open
squares). The DLA measurements in Figure~\ref{undern} are from
Centuri\'{o}n et al. (2003, and references therein) with the
exceptions and additions summarized by Jenkins et
al. (2004).\footnote{Jenkins et al. (2004) have carefully screened the
high$-z$ DLA data, particularly to check for abundance estimates that
may be underestimated due to saturation, and substituted unsaturated
$\alpha-$ element abundances when available.}  The \ion{H}{2}-region
emission-line data in Figure~\ref{undern} were compiled by Pilyugin et
al. (2003) and converted to abundances using the ``P-method'', which
appears to be in better agreement with \ion{H}{2}-region abundances
with electron temperature measurements (e.g. Bresolin, Garnett, \&
Kennicutt 2004) than some of the more commonly used ``R23''
calibrations.\footnote{See Kobulnicky \& Kewley (2004) for a detailed
discussion and comparison of R23 calibrations from the literature.
Kobulnicky \& Kewley note that the P-method might not be
well-calibrated in the high-metallcity regime, but this does not
affect the main conclusions that we draw from Figure~\ref{undern}
because the most relevant comparison is to the low-metallicity H~II
region measurements. Use of other R23 calibrations would increase the
$\alpha$ abundances by a few tenths of a dex for the
higher-metallicity H~II regions.}  For comparison, figures analogous
to Figure~\ref{undern} but using the traditional R23 calibration can
be found in Pettini et al. (2002) and Prochaska et al. (2002b). From
Figure~\ref{undern}, we see that the N underabundance in the Virgo
sub-DLA toward PG1216+069 is similar to that seen in nearby giant
\ion{H}{2} regions and distant DLAs, but the absolute metallicity of
the sub-DLA is lower than the metallicities of the vast majority of
the nearby galaxies.

For the most part, the second (weaker) component at $v = 28$ \kms\
offers fewer constraints and noiser measurements. We note that the
column density ratios agree in both components to within the
measurement uncertainties.  The \ion{C}{2} column density, however, is
more reliably measured in the weaker component [$N$(\ion{C}{2}) is
uncertain in the stronger component due to saturation], and the carbon
and oxygen columns indicate that [C/O] $\approx$ 0 if ionization
corrections are neglected. This is somewhat surprising because
observations of metal-poor Galactic halo stars by Akerman et
al. (2004) indicate that [C/O] $\approx -0.5$ when [O/H] $\lesssim -1$
(see their Figure 5).  Akerman et al. also report a tentative
indication that [C/O] may return to near-solar values in the
lowest-metallicity stars. The weaker component of the PG1216+069
sub-DLA may therefore have an abundance pattern more similar to the
most metal-poor stars in the Akerman sample. However, [C/O] in the
sub-DLA is sensitive to ionization corrections (note how the C/O ratio
changes as $U$ increases in Figure~\ref{photmodel}). The uncertain
ionization correction and measurement uncertainties of the current
data allow a substantially lower value of [C/O] in the PG1216+069
sub-DLA.

\subsection{Molecular Hydrogen}

We have investigated whether or not absorption features from molecular
hydrogen can be seen in the FUSE spectrum of PG1216+069 at the sub-DLA
redshift.  Under the most probable conditions for rotational
excitation ($50 \lesssim T_{\rm rot}\lesssim 1000\,$K), we expect the
strongest H$_{2}$ features to be those from the $J=1$ level.  At the
positions of three strong transitions from this level, the Werner
0$-$0 R(1) and Q(1) lines plus the Lyman 7$-$0~R(1) line, we measured
an average for the equivalent widths equal to $W_{\rm r}=40\,$m\AA\
with a $1\sigma$ uncertainty of 30$\,$m\AA, which does not constitute
a real detection.  We may derive a $3\sigma$ upper limit for $N({\rm
H}_2)$ in the $J=1$ level by evaluating the column density on the
damping part of the curve of growth for $W_{\rm r}=130\,$m\AA\ for any
of the three lines, all of which have nearly the same wavelengths,
$f-$values (Abgrall \& Roueff 1989) and lifetimes (Morton \&
Dinerstein 1976).  Our upper limit is $N({\rm H}_2)= 7\times
10^{17}{\rm cm}^{-2}$ in the $J = 1$ level. If we assume the maximum
H$_{2}$ rotational temperature observed in the Milky Way, $T_{\rm rot}
\approx 1000$ K, then the upper limit on H$_{2}$ in all rotational
levels is roughly $3\times N({\rm H}_{2},J=1)$ (Jenkins et al. 2004
and references therein). Therefore our upper limit for the fraction of
hydrogen atoms in molecular form $f({\rm H}_2)\equiv 2N({\rm
H}_2)/[2N({\rm H}_2)+N($\ion{H}{1}$)]< 0.17$.  This limit is above
values found for high$-z$ DLAs (Curran et al. 2004 and references
therein) and the Magellanic Clouds (Tumlinson et al. 2002), but is
below measurements toward some moderately reddened stars in our Galaxy
(Savage et al. 1977).  However, when the \ion{H}{1} column density of
the PG1216+069 sub-DLA is taken into account, we see from Figure 8 in
Tumlinson et al. (2002) that a rather low $f({\rm H}_2)$ value would
be expected even in Milky Way sight lines. Also, since H$_{2}$ forms
primarily on the surface of dust grains, a low H$_{2}$ fraction is
expected in the Virgo sub-DLA given the low metallicity and dust
content derived in the previous sections.

\section{Physical Conditions and Neutral Gas Content\label{physcon}}

In addition to abundance constraints, the ionization models provide
information on physical conditions of the gas such as the particle
number density, temperature, thermal pressure, and neutral fraction of
the gas. The gas ion fractions and therefore the neutral gas content
of the absorber are generally well-constrained by the models. Other
physical quantities suffer greater uncertainties. For example, in
principle $n_{\rm H}$ is directly related to the ionization parameter
$U$, but there is a caveat: since $U = n_{\gamma}/n_{\rm H}$, the
derived $n_{\rm H}$ depends on the intensity taken for the ionizing
radiation field. For the model shown in Figure~\ref{photmodel}, we
have assumed $J_{\nu} = 10^{-23}$ ergs cm$^{-2}$ s$^{-1}$ Hz$^{-1}$
sr$^{-1}$, a value in good agreement with current observational bounds
on the UV background from QSOs at $z \approx$ 0. However, $J_{\nu}$ is
still only loosely bracketed by observations (see, e.g., Shull et
al. 1999; Dav\'{e} \& Tripp 2001; Weymann et al. 2001). Moreover,
given the high $N$(\ion{H}{1}) of the PG1216+069 sub-DLA, it is
possible that stars are located in or near the gas.  We have
constructed photoionization models for the Virgo sub-DLA assuming that
nearby hot stars contribute to the ionizing radiation in addition to
the flux from background quasars, and we reach generally similar
conclusions regarding the ionization corrections and neutral gas
fraction based on the stars+QSOs model. However, adding stars
increases $n_{\gamma}$, and therefore a stars+QSOs model at the same
value of $U$ has to have a higher $n_{\rm H}$ than the analogous
QSOs-only model. For this reason, the $n_{\rm H}$ constraints from the
QSOs-only model are most conservatively treated as {\it lower
limits}. In turn, the implied gas pressure is also a lower limit, and
the line-of-sight absorber size is an upper limit. The mean gas
temperature also depends on the model parameters, but less
sensitively.

\begin{figure}
\epsscale{1.0}
\plotone{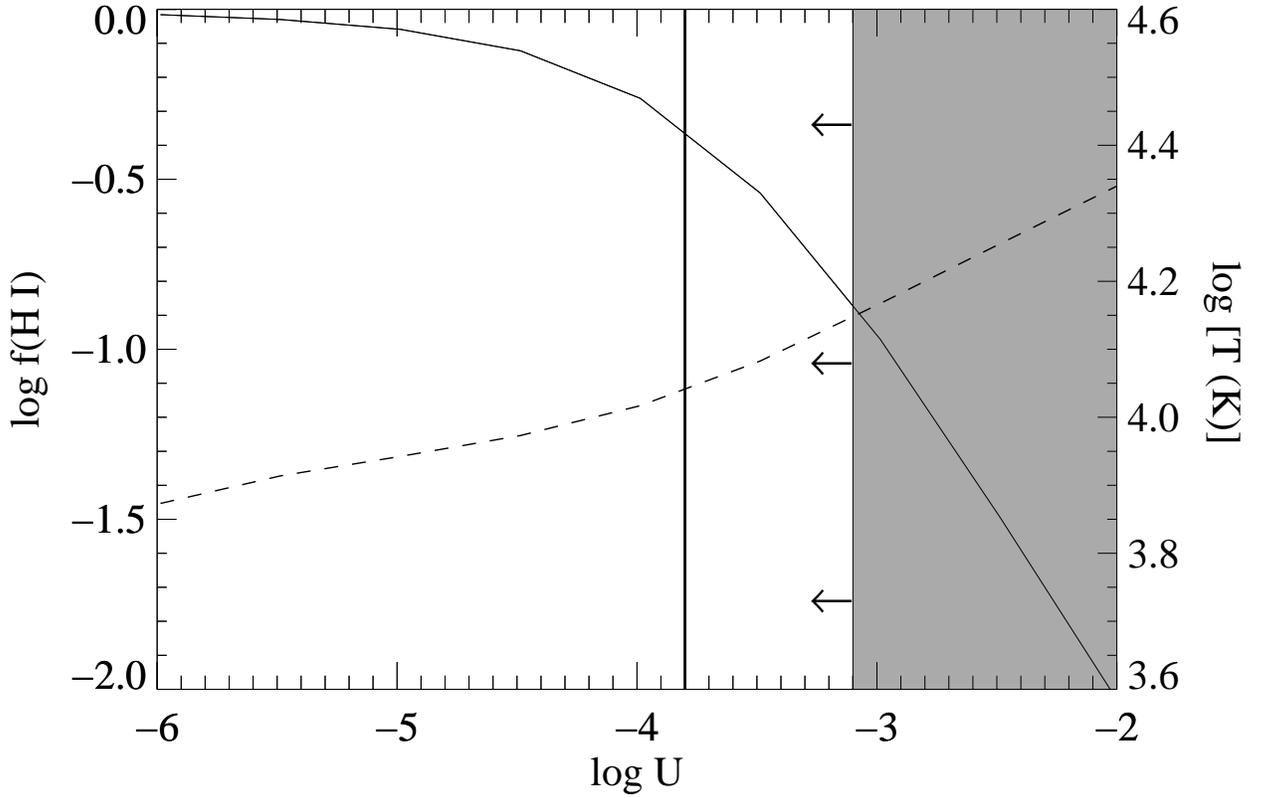}
\caption[]{Neutral-gas content and mean temperature of the sub-DLA
according to the photoionization model shown in
Figure~\ref{photmodel}. The \ion{H}{1} ion fraction is plotted with a
solid line using the scale on the left axis, and the temperature is
indicated with a dashed line using the scale on the right axis; both
quantities are plotted vs. the ionization parameter $U$. The upper
limits on \ion{C}{4} and \ion{Si}{4} require log $U \lesssim -3.1$, as
shown by the gray region. The thick vertical line is drawn at the
value of $U$ that provides the best fit to the observed column
densities (see Figure~\ref{photmodel}).\label{neutralfrac}}
\end{figure}

With this caveat, we show in Figures~\ref{photmodel} and
\ref{neutralfrac} some of the physical conditions implied by the
photoionization modeling. The top axis of Figure~\ref{photmodel} shows
the particle density for the QSOs-only ionizing radiation field for
the assumed value of $J_{\nu}$ at 1Ryd; we treat these densities as
lower limits. Figure~\ref{neutralfrac} presents the \ion{H}{1} ion
fraction (left axis) and mean gas temperature (right axis) for the
same model as a function of log $U$. The upper limit on
$N$(\ion{Si}{4}) requires log $n_{\rm H} \geq -3.4$, log
$f$(\ion{H}{1}) $\geq -0.9$, $T \leq$ 14,000 K, and thermal pressure
$P/k \geq$ 10 K cm$^{-3}$.  In this model, the line-of-sight absorber
size is less than $\sim$100 kpc. We see from Figure~\ref{neutralfrac}
that even though high ions such as \ion{Si}{4} and \ion{C}{4} are not
detected in the Virgo sub-DLA, it could still contain a substantial
amount of ionized gas. At the best-fitting ionization parameter (log
$U \approx -3.8$, shown with a thick vertical line in
Figure~\ref{neutralfrac}), $\sim$60\% of the hydrogen is
ionized. While the particle number density and absorber size are
uncertain, it is interesting to note that the gas could be
self-gravitating with the physical conditions indicated by the model
in Figure~\ref{photmodel} at log $U \approx -3.8$, assuming the
baryon-to-dark matter ratio $\Omega_{b}/\Omega_{m} \approx$ 0.15 (see,
e.g., Schaye 2001). Other confinement options (e.g., pressure
confinement) remain viable, but gravitational confinement is not ruled
out.

\section{Absorber Environment\label{abenv}}

\begin{figure}
\plotone{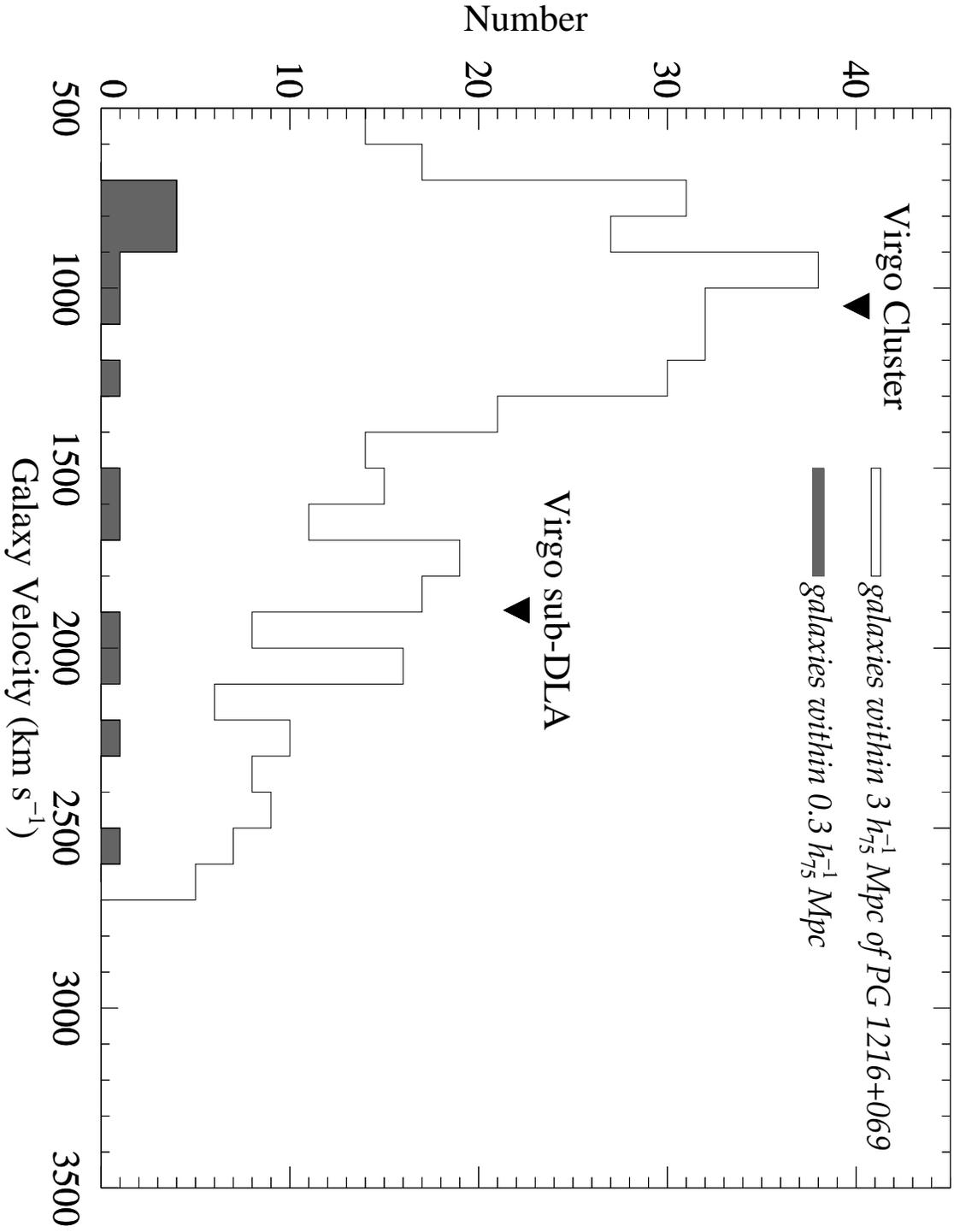}
\caption[]{Number of RC3 galaxies within projected distances of
$3h_{75}^{-1}$ Mpc (open histogram) and $300h_{75}^{-1}$ kpc (filled
histogram) of the PG1216+069 sight line, plotted vs. redshift in 100
\kms\ bins. The redshift of the sub-DLA discussed in this paper is
indicated.\label{galdensity}}
\end{figure}

As noted in \S1 and shown in Figure~\ref{xraymap}, the sight line to
PG1216+069 pierces the outskirts of the Virgo cluster and the X-ray
bright NGC4261 galaxy group (Davis et al. 1995). Although the sight
line does not penetrate the Virgo cluster 6$^{\circ}$ core, it does
pass within the cluster's maximum angle of influence according to the
model of Tully \& Shaya (1984). The QSO is also in the general
direction of the Virgo W and W$^{\prime}$ structures south of the
cluster proper (e.g., de Vaucouleurs 1961; Binggeli et
al. 1993). Compared to most low$-z$ \lya absorbers, the PG1216+069
sub-DLA is certainly located in a region of relatively high galaxy
density. To show this, we plot in Figure~\ref{galdensity} the number
of galaxies from the RC3 catalog (de Vaucouleurs et al. 1991) within
projected distances of 3$h_{75}^{-1}$ Mpc and 300$h_{75}^{-1}$ kpc
from the PG1216+069 sight line. Comparing Figure~\ref{galdensity} to
the analogous figures in Bowen, Pettini, \& Blades (2002), we see that
on the 3 Mpc scale, the galaxy density is substantially larger than
the galaxy density in the vicinity of a typical low$-z$ \lya cloud.
From Figure~\ref{galdensity} we see that the PG1216+069 sub-DLA is at
a somewhat higher velocity than the bulk of the galaxies in the Virgo
cluster proper; the sub-DLA is closer in velocity to the Virgo W
structure.

Given the high $N$(\ion{H}{1}) of the absorber, some models would
predict that the absorption arises in the bound ISM of a Virgo galaxy,
which one might expect to find near the sight line. However, the
nearest $L*$ galaxy near $z \sim$ 0.00632 is NGC4260 (Bowen et
al. 1996; Impey et al. 1999), which has a surprisingly large impact
parameter $\rho$ of 246 $h_{75}^{-1}$ kpc. As noted by Impey et
al. (1999), VCC297 is also near the sub-DLA redshift and has a smaller
projected distance, $\rho = 86 h_{75}^{-1}$ kpc. The luminosity of
VCC297 is less than $0.25 L*$.  NGC4260 is, in fact, the nearest
galaxy to the PG1216+069 sight line that Davis et al. (1995) identify
within the diffuse X-ray emitting gas in the NGC4261 group.  The RC3
redshift of NGC4260 is 1886$\pm$61 \kms , which is the same as the
sub-DLA redshift within the 1$\sigma$ uncertainty. The redshifts of
the other galaxies in this group listed by Davis et al. range from
2210 \kms\ (NGC4261) to 2685 \kms\ (NGC4281), i.e., within 300-800
\kms\ of the absorber redshift. NGC4261 is a giant elliptical galaxy
located near the peak of the X-ray emission; this is likely the center
of mass of the group.  If the PG1216+069 sub-DLA is within the NGC4261
group and the redshift of NGC4261 reflects the group distance, then
the absorber impact parameter to the group $\sim$400 $h_{75}^{-1}$
kpc.

\begin{figure}
\plotone{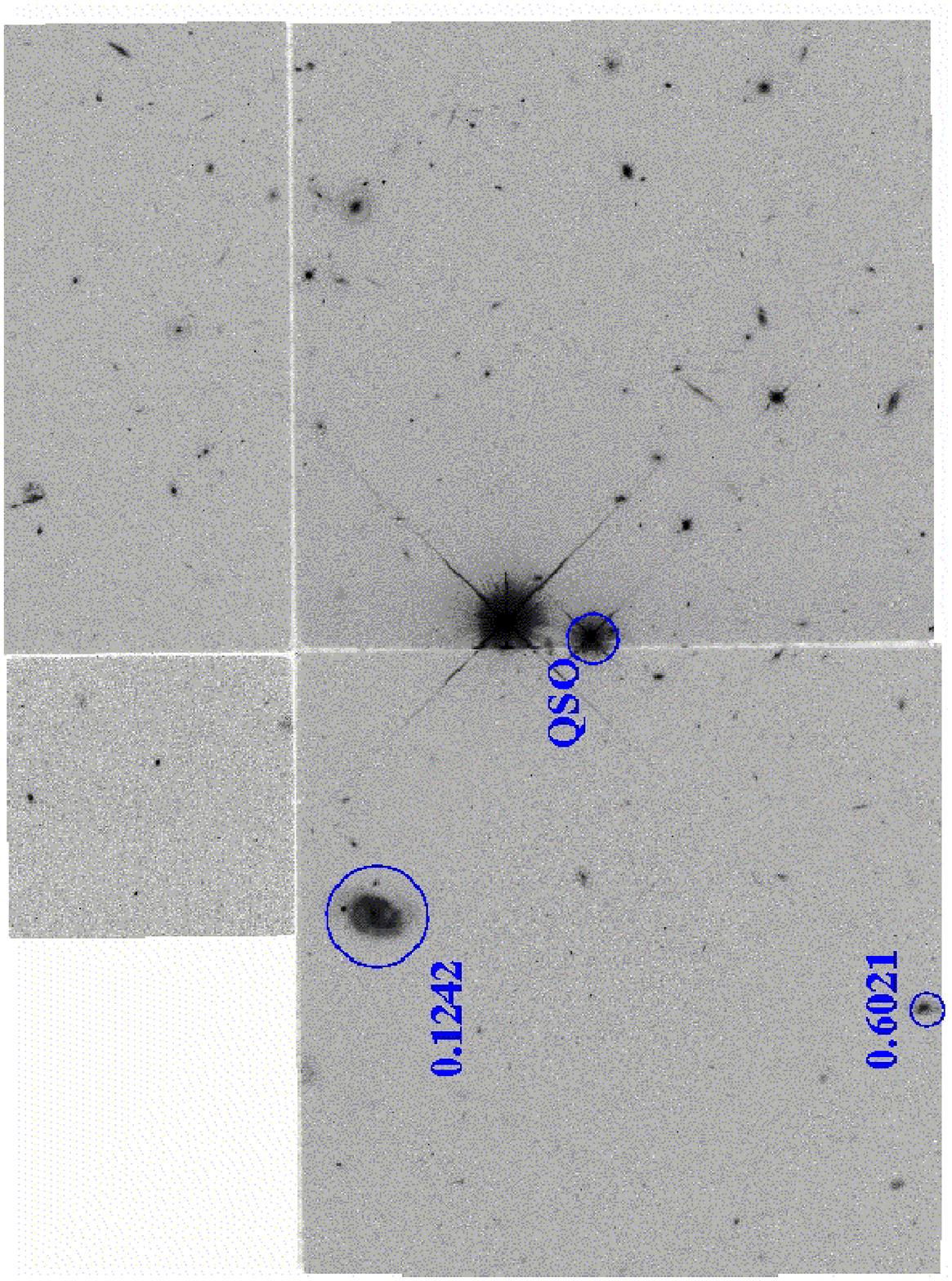}
\caption[]{Portion of an image of the field surrounding PG1216+069
obtained with the WFPC2 camera on board {\it HST}. All galaxies within
the image that have spectroscopically-measured redshifts are marked
with the measured redshift.  The QSO is also circled. The image spans
2.7' in the x direction.\label{wfpcimage}}
\end{figure}

\begin{figure}
\plotone{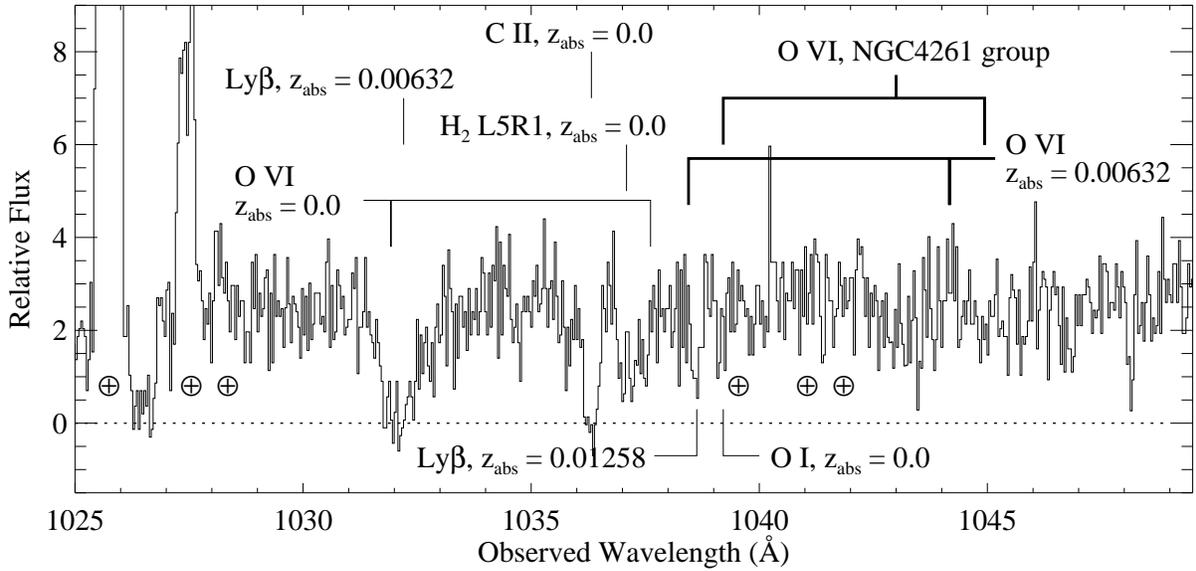}
\caption[]{Portion of the {\it FUSE} spectrum covering the expected
wavelengths of the \ion{O}{6} $\lambda \lambda$1031.93, 1037.62
doublet at the redshift of the Virgo sub-DLA and the NGC4261
group. Well-detected absorption lines are labeled, and terrestrial
airglow emission line wavelengths are also marked ($\earth$). The
stronger line of the \ion{O}{6} doublet at the Virgo/NGC4261 redshift
falls in a region complicated by airglow emission, Galactic \ion{O}{1}
$\lambda$1039.23 absorption (which appears to be filled in by airglow
emission), and Lyman $\beta$ absorption from \zabs\ = 0.01258. No
significant absorption is evident at the wavelength of the weaker
\ion{O}{6} line at Virgo redshifts.\label{fuseo6spec}}
\end{figure}

Chen et al. (2001) have obtained high-resolution {\it HST}/WFPC2
imaging and spectroscopic galaxy redshifts in order to study the
relationship between the galaxies and \lya clouds in the direction of
PG1216+069, and they have not found any close galaxies at Virgo
redshifts either. Figure~\ref{wfpcimage} shows a portion of a WFPC2
image of the QSO with the redshifts from Chen et al. labeled. While a
number of galaxies with unknown redshifts are readily apparent in
Figure~\ref{wfpcimage}, based on their angular sizes, these galaxies
are likely to be distant background galaxies beyond Virgo. The
galaxies with unknown redshifts are also quite faint; apart from the
extended spiral at $z$ = 0.1242, most of the galaxies near the sight
line in Figure~\ref{wfpcimage} have $m \gg$ 21 in the WFPC2 F702W
filter. We note that a bright foreground star is present near
PG1216+069, and while unlikely, a Virgo galaxy could be masked by this
foreground object.

\section{Warm-Hot Gas in the NGC4261 Galaxy Group\label{whimsec}}

The proximity of the sight line to the X-ray bright NGC4261 group
provides an opportunity to test ideas regarding the location of the
``missing baryons'' at the present epoch. As has been widely discussed
by various authors (e.g., Persic \& Salucci 1992; Fukugita, Hogan, \&
Peebles 1998) the combined masses in well-observed baryonic components
of the nearby universe do not add up to the expected mass in baryons
predicted by D/H measurements and cosmic microwave background
observations (e.g., Spergel et al. 2003; Sembach et al. 2004b), the
so-called missing-baryons problem.  Hydrodynamic simulations of
cosmological structure evolution suggest that these missing baryons
are located in ``warm-hot'' shock-heated gas at $10^{5} - 10^{7}$ K in
regions of the intergalactic medium at modest overdensities (e.g., Cen
\& Ostriker et al. 1999; Dav\'{e} et al. 2001). In contrast, Fukugita
et al. (1998) have hypothesized that the missing baryons are
predominantly found in similarly-hot gas, but in higher overdensity
regions near galaxy groups. In either case, the \ion{O}{6} doublet
provides a sensitive means to search for the warm-hot gas since the
\ion{O}{6} ion fraction peaks at $T \sim 10^{5.5}$ K.  Observations of
low-z QSOs have frequently revealed redshifted \ion{O}{6} absorption
lines that are clearly correlated with galaxies (e.g., Tripp \& Savage
2000; Chen \& Prochaska 2000; Tripp et al. 2001; Savage et al. 2002;
Sembach et al. 2004a), but so far these studies have not had sufficient
information regarding the nearby galaxies to establish if they form a
bound group with the overdensity expected in the Fukugita et
al. hypothesis or if they are situated in lower overdensity
unvirialized environments as expected from the cosmological
simulations.

A few preliminary searches for \ion{O}{6} associated with well-defined
galaxy groups have been attempted (e.g., Wakker et al. 2003; Pisano et
al. 2004), but no definitive detections of \ion{O}{6} have yet been
reported.  The PG1216+069 sight line through the NGC4261 group is
appealing for this purpose because the diffuse X-ray emission
establishes that the group certainly contains intragroup gas, at least
in its inner regions, and because the sight line impact parameter is
relatively small (compared to available QSO/group pairs bright enough
for high-resolution UV spectroscopy with current facilities).
Figure~\ref{fuseo6spec} shows the portion of the {\it FUSE} spectrum
of PG1216+069 covering the \ion{O}{6} doublet at the redshifts of the
NGC4261 group (the expected wavelengths of \ion{O}{6} at the redshift
of the sub-DLA are also marked). No \ion{O}{6} absorption is evident
at the NGC4261 or sub-DLA redshifts. Unfortunately, the stronger
\ion{O}{6} $\lambda$1031.93 line falls in a messy region affected by
the Galactic \ion{O}{1} $\lambda$1039.23 absorption, a Ly$\beta$ line
at \zabs\ = 0.01258, and terrestrial \ion{O}{1} airglow emission lines
(Feldman et al. 2001).\footnote{Based on the strength of the
\ion{O}{1} $\lambda$1302.17 transition in the STIS spectrum,
\ion{O}{1} $\lambda$1039.23 should be detectable in the {\it FUSE}
data. Evidently the 1039.23 line is largely filled in by terrestrial
\ion{O}{1} $\lambda$1039.23 emission.} Consequently, we must rely on
the weaker line to place upper limits on the \ion{O}{6} column
density.  Integrating over the velocity range of the detected
low-ionization lines in the sub-DLA, we obtain a 3$\sigma$ upper limit
of log $N$(\ion{O}{6}) $\leq$ 14.26; limits at the slightly higher
redshift of the NGC4261 group are similar assuming a comparable
velocity integration range. 

This constraint is limited by the modest signal-to-noise ratio of the
data. Nevertheless, the upper limit is comparable to the column
densities of the stronger \ion{O}{6} absorbers detected in other
directions (e.g., Tripp et al. 2000a; Chen \& Prochaska 2000; Savage
et al. 2002; Howk, Prochaska, \& Chen 2004).  Given the detection of
X-ray emitting gas with ROSAT (Davis et al. 1995), it seems possible
that the intragroup gas is too hot to show \ion{O}{6} (i.e., the
oxygen is predominantly in higher ionization stages). However, in this
case the sub-DLA might be expected to show, e.g., \ion{C}{4} from the
interface between the hot X-ray gas and the low-ionization gas.

\section{Discussion\label{disec}}

The absorber properties derived above for the PG1216+069 sub-DLA are,
in some regards, surprising. The gas metallicity is extremely low
given the location of the gas in a region of relatively high galaxy
density near the Virgo cluster and an X-ray bright galaxy group.  In
fact, the absolute abundance of this sub-DLA is comparable to the
lowest {\it gas-phase} metallicities known in the nearby universe (see
Figure~\ref{undern}). For example, the blue compact dwarf (BCD)
galaxies I Zw 18 and SBS0335-052 are among the most metal-poor
galaxies so far found at low redshifts. Analysis of \ion{H}{2} region
emission lines indicates that these BCD galaxies have gas-phase
metallicities $Z \approx 0.02 Z_{\odot}$ (Izotov et
al. 1999). Measurement of absorption lines in I Zw 18 yield similar
low metallicities; possibly the abundances in the diffuse \ion{H}{1}
regions probed by the absorption lines are a few tenths of a dex lower
than the \ion{H}{2} region abundances (Aloisi et al. 2003; Lecavelier
des Etangs et al. 2004).  Even lower metallcities have been observed
in high$-z$ QSO absorption systems and some Galactic stars, but
high$-z$ absorbers and low-metallcity stars both probe earlier epochs
in the history of the universe; the DLA discussed in this paper
provides information about the {\it current} metallicity of the
absorbing entity.

But what is the nature of the absorbing entity?  The absorber has
similarites to some well-studied low$-z$ objects.  The PG1216+069
sub-DLA is reminiscent of the BCD galaxy I Zw 18 in many ways.  As
already noted, both objects have extremely low gas-phase
metallicities.  Like the PG1216+069 sub-DLA, I Zw 18 contains a lot of
neutral gas (van Zee et al. 1998b; Aloisi et al. 2003; Lecavelier des
Etangs et al. 2004). However, the most intriguing similarities of the
sub-DLA and the BCD galaxy are the relative metal abundances. Aloisi
et al. (2003) find that nitrogen is highly underabundant in I Zw 18,
and yet they find no clear indication of $\alpha-$element
overabundances.  On the contrary, they report that the
$\alpha-$elements O and Si are marginally {\it underabundant} with
respect to iron. The abundance pattern in I Zw 18 is apparently
similar to the abundance pattern that we find in the Virgo sub-DLA
(see \S~\ref{abund}). The apparent lack of an identified counterpart
for the PG1216+069 sub-DLA could be used to argue against the
hypothesis that the sub-DLA is somehow related to a BCD galaxy. One
might expect to see evidence of a I Zw 18-like galaxy in the image
shown in Figure~\ref{wfpcimage}. However, the QSO itself as well as
the bright foreground star could mask a faint galaxy close to the
sight line.

The high-velocity clouds (HVCs) in the vicinity of the Milky Way are
also similar to the PG1216+069 sub-DLA in several respects.  The HVCs
have comparable \ion{H}{1} column densities (Wakker \& van Woerden
1997; Wakker 2001), and when HVC metallicities have been measured,
they have showed similar abundance patterns. For example, HVC Complex
C has a sub-solar absolute abundance, [O/H] $\approx -0.8$, it shows a
significant nitrogen underabundance, and Complex C iron abundances
indicate that there is little or no dust depletion (Murphy et
al. 2000; Richter et al. 2001; Gibson et al. 2001; Collins et al 2003;
Tripp et al. 2003; Sembach et al. 2004b). In all of these respects, the
PG1216+069 Virgo sub-DLA is analogous to Complex C, although the
sub-DLA has a substantially lower overall metallicity than Complex C.
Some redshifted gas clouds analogous to the Milky Way HVCs have been
detected in 21 cm emission studies, e.g., the Leo ring in the M96
group (Schneider 1989) or the large \ion{H}{1} cloud near 3C 273
(Giovanelli \& Haynes 1989).  Usually these 21 cm clouds have been
shown to be closely associated with some type of galaxy, albeit faint
in some cases (e.g., Salzer et al. 1991).

It is worth noting that a lack of $\alpha-$element enhancement (with
respect to Fe) at low absolute metallicity has been also been observed
in stars in several nearby dwarf spheroidal (dSph) galaxies (Shetrone,
C\^{o}t\'{e}, \& Sargent 2001; Tolstoy et al. 2003). Dwarf spheroidal
galaxies generally have only small amounts of associated neutral
interstellar gas, at least in the main galaxy where the stars are
found (e.g., Carignan 1999).  Nevertheless, there may be a connection
between high-$N$(\ion{H}{1}) absorbers and dSph's. Consider the
Sculptor dSph galaxy. In this galaxy, 21cm observations show little
emission centered on the stars in the galaxy, but two lobes of 21cm
emission are detected at offsets of $15-20$' from the optical
(stellar) galaxy (Carignan et al. 1998). The symmetric morphology of
the 21cm emission is suggestive of a bipolar outflow. Removal of
interstellar gas from dSph's, either by supernovae-driven winds or by
dynamical processes (tidal or ram-pressure stripping), would help to
explain the paradoxical indications of extended periods of
star-formation in dSph galaxies, which is difficult to reconcile with
their current lack of interstellar gas (Blitz \& Robishaw 2000;
Grebel, Gallagher, \& Harbeck 2003).  Once the gas has been removed
from the dSph galaxy, it could then give rise to a DLA.  If a QSO
sight line happened to pass through one of the 21cm emission lobes
near Sculptor, for example, the resulting absorber properties would
likely be very similar to the PG1216+069 sub-DLA.

Bowen et al. (1997) have searched for absorption lines in the vicinity
of the dSph galaxy Leo I using three background QSOs/AGNs at projected
distances ranging from 2-8 kpc. One of their spectra has insufficient
signal-to-noise to place interesting limits, but the spectra of the
other two QSOs show no absorption at the velocity of Leo I, and Bowen
et al. argue that $N$(\ion{H}{1}) $\lesssim 10^{17}$ cm$^{-2}$ within
2-4 kpc of that dSph. Leo I therefore presents an interesting contrast
to the Sculptor dSph, and the Bowen et al. study argues against the
hypothesis that dSph's are a source of high-$N$(\ion{H}{1}) absorption
systems. There are various possible explanations for the differences
between Sculptor, which shows high-$N$(\ion{H}{1}) in its immediate
vicinity, and Leo I, which apparently does not. Sculptor is one of the
nearest of the satellite galaxies of the Milky Way; Leo I is at least
twice as far away.  Perhaps gas is only stripped when the dSph's pass
close to the Milky Way (Blitz \& Robishaw 2000; Grebel et al. 2003)
and then dissapates relatively quickly (see Bowen et
al. 1997). However, there are no known luminous, large galaxies near
the PG1216+069 sight line, so if this is the correct explanation for
the difference between Sculptor and Leo I, it then seems unlikely that
the PG1216+069 sub-DLA is related to a dSph.  Another possibility is
that Sculptor and its associated clouds are related to the Magellanic
Stream, which is located in the Sculptor direction and has similar
velocities (Putman et al. 2003).

A final possibility is that the PG1216+069 sub-DLA arises in a small,
ancient dark-matter minihalo that formed before the epoch of
reionization and subsequently evolved largely undistrubed. Abel \& Mo
(1998) have argued that minihalos formed before reionization could
accumulate and retain substantial quantities of \ion{H}{1}, and moreover
their densities would likely be high enough so that the objects would
remain neutral after the substantial increase in background UV flux at
the reionization epoch.  Minihalos formed after reionization would, in
contrast, be more highly ionized, hotter, and less able to accrete and
retain gas (e.g., Bullock, Kravtsov, \& Weinberg 2000). Maller et
al. (2003) have recently revisited this idea, and they conclude that
while minihalos are unlikely to give rise to Lyman limit absorbers,
they could be a significant source of sub-DLA systems.  

The simple kinematical structure of the PG1216+069 sub-DLA is
consistent with expectations for minihalos, which would have circular
velocities $v_{c} \lesssim$ 30 \kms . Kepner et al. (1999) have
modeled the absorption-line signatures of photoionized gas in
dark-matter minihalos, and their column density predictions, including
the \ion{H}{1} column density, the upper limit on molecular hydrogen,
and the low ion/high ion column density ratios of the metals, are
consistent with the observed properties of the Virgo sub-DLA if the
sight line intercepts the minihalo at $\rho \sim$ 1 kpc (see Figures 2
and 5 in Kepner et al).  Bearing in mind the caveats of
\S~\ref{physcon}, it is interesting to note that the line-of-sight
thickness of the best-fitting CLOUDY model shown in
Figure~\ref{photmodel} would be consistent with the size implied by
the Kepner et al. model if log $U \lesssim -4.0$.  Given that there
could be some starlight that contributes to the ionizing flux (which
makes the absorber thickness smaller), the Kepner et al. size can be
reconciled with the photoionization models for the PG1216+069 sub-DLA.
The sub-DLA metallicity is larger than metallicities typically derived
for diffuse Ly$\alpha$ forest gas at high redshifts, so this
hypothesis would require some self-enrichment from internal stars in
the minihalo.  Very low-level star formation and self enrichment are
plausible, e.g., triggered by occassional perturbations of the gas,
but one weakness of this idea is that a small number of supernovae
could blow away the gas in the minihalo.

\section{Summary\label{sumsec}}

Using a new high-resolution observation of PG1216+069 obtained with
the E140M echelle mode of STIS, we have identified a nearby sub-damped
\lya absorber at \zabs\ = 0.00632 with log $N$(\ion{H}{1}) =
19.32$\pm$0.03.  This absorption system has several remarkable
properties. First, the overall metal abundance is surprisingly low; we
find [O/H] = $-1.60^{+0.09}_{-0.11}$ and $-1.77 \leq$ [Si/H] $\leq
-1.35$.  These are among the lowest metallicity measurements {\it in
the gas phase} known in the nearby universe (stars can have much lower
abundances, but low stellar abundances are fossils that mainly provide
information about earlier epochs). Second, the iron abundance
indicates that this absorber harbors very little dust.  Third,
nitrogen is significantly underabundant, which indicates that the gas
is ``chemically young''.  However, comparison of the oxygen, silicon,
and iron abundances shows no significant evidence of $\alpha-$element
overabundances.  Fourth, although the gas is situated in the outer
reaches of the Virgo galaxy cluster and the NGC4261 galaxy group,
there are no known luminous galaxies within the projected distance
that might be expected for such a high-$N$(\ion{H}{1}) absorber. The
nearest $L*$ galaxy is NGC4260 at a projected distance of 246
$h_{75}^{-1}$ kpc, and even the closest known sub-luminous galaxy is
at a substantial projected distance, $\rho = 86 h_{75}^{-1}$ kpc. We
place limits on \ion{O}{6} absorption associated with the NGC4261
galaxy group, which is known to have hot intragroup gas from detection
for diffuse X-ray emission.

The abundance patterns in the sub-DLA are reminiscent of abundances in
nearby dwarf galaxies and of gas-phase abundances in high-velocity
clouds near the Milky Way.  We discuss the possibility that this
absorber is related to a dwarf galaxy and/or a high-velocity cloud in
the outer region of the Virgo galaxy cluster. It is also possible that
the PG1216+069 sub-DLA arises in a small dark-matter halo. In this
case, it is likely that the halo formed and accumlated gas in the
early universe before reionization occurred, and since that time has
only undergone meager self-enrichment from internal star
formation. Several future measurements would be helpful for
understanding the nature of this absorber. Better measurements of the
Fe abundance would be valuable for learning about relative abundance
patterns in the sub-DLA; in addition, the strong \ion{Fe}{2} lines and
higher signal-to-noise achievable with the far-UV mode of STIS would
enable detection of weaker absorption components. We have shown that
weak components will not affect the derived total abundances much, but
the {\it kinematics} implied by the presence (or absence) of higher
velocity features would be useful for testing the small dark-matter
halo interpretation.  A minihalo should have a low circular velocity
and simple kinematics, and high-velocity weak components are not
expected in this model. A deep 21 cm emission map with high spatial
resolution at Virgo redshifts would test several of the possible
interpretations we have presented for the sub-DLA. For example, if the
sub-DLA is part of a tidally-stripped gas structure or is analogous to
the \ion{H}{1} ring in the M96 group or the Giovanelli \& Haynes
cloud, this would be revealed by deep 21 cm observations. Moreover,
such structures would have clearly different morphologies from a
minihalo or compact high-velocity cloud. Narrow-band H$\alpha$ imaging
could also provide useful clues about the nature of the sub-DLA, e.g.,
to search for evidence of a faint star-forming galaxy near the sight
line. Regardless of the detailed nature of the absorber, this
observation unambiguously shows that relatively primitive gas that has
undergone little chemical enrichment (compared to metallicities in
high-redshift gases) is still found at the present epoch.

\acknowledgements 

This paper benefitted from valuable discussions with Ari Maller, Max
Pettini, Blair Savage, Ken Sembach, and Jason Tumlinson.  We also
appreciate useful comments from the referee.  The STIS observations
were obtained under the auspices of {\it HST} program 9184, with
financial support through NASA grant HST-GO-9184.08-A from the Space
Telescope Science Institute.  We also appreciate and acknowledge
support from NASA LTSA grant NNG04GG73G. The {\it FUSE} data were
obtained by the PI team of the NASA-CNES-CSA {\it FUSE} project, which
is operated by Johns Hopkins University with financial support through
NASA contract NAS 5-32985. This research has made use of the NASA/IPAC
Extragalactic Database (NED), which is operated by the Jet Propulsion
Laboratory, California Institute of Technology, under contract with
NASA.

\appendix

\section{STIS Wavelength Calibration Errors}
\begin{figure}
\plotone{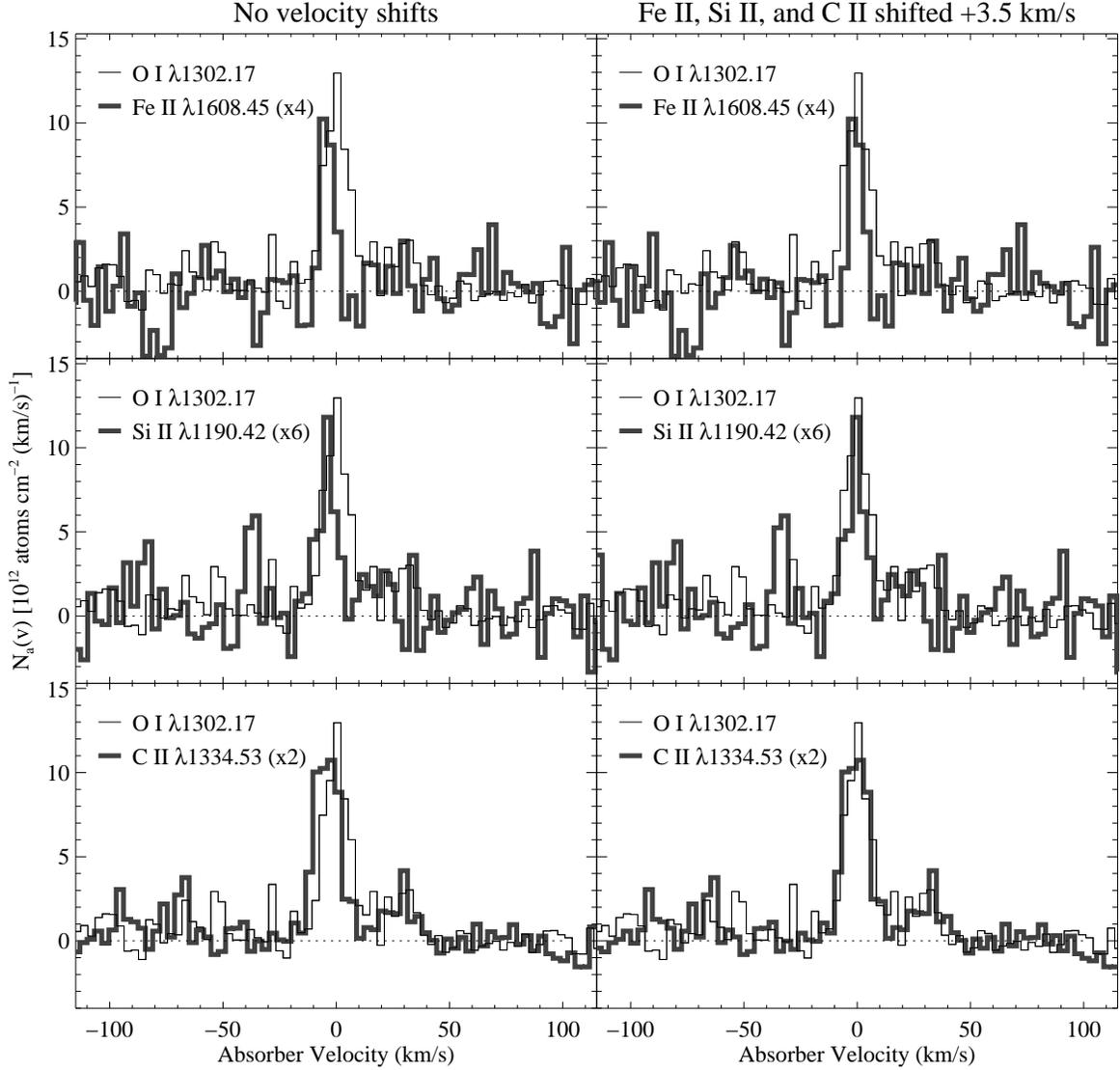}
\caption[]{Comparison of the apparent column density profile of
\ion{O}{1} $\lambda$1302.17 to the \nav\ profiles of \ion{Fe}{2}
$\lambda$1608.45 (top), \ion{Si}{2} $\lambda$1190.42 (middle), and
\ion{C}{2} $\lambda$1334.53 (bottom). The left panels show the \nav\
profiles with no shifts applied to the velocity scale; significant
differences are evident in the centroids of the various lines. The
right panels show the same comparisons, but with the \ion{Fe}{2},
\ion{Si}{2}, and \ion{C}{2} profiles shifted by +3.5 \kms , or roughly
one pixel. For purposes of comparison, the \ion{Fe}{2}, \ion{Si}{2},
and \ion{C}{2} \nav\ profiles have been scaled up by factors of 4, 6,
and 2, respectively.\label{oshift}}
\end{figure}

\begin{figure}
\plotone{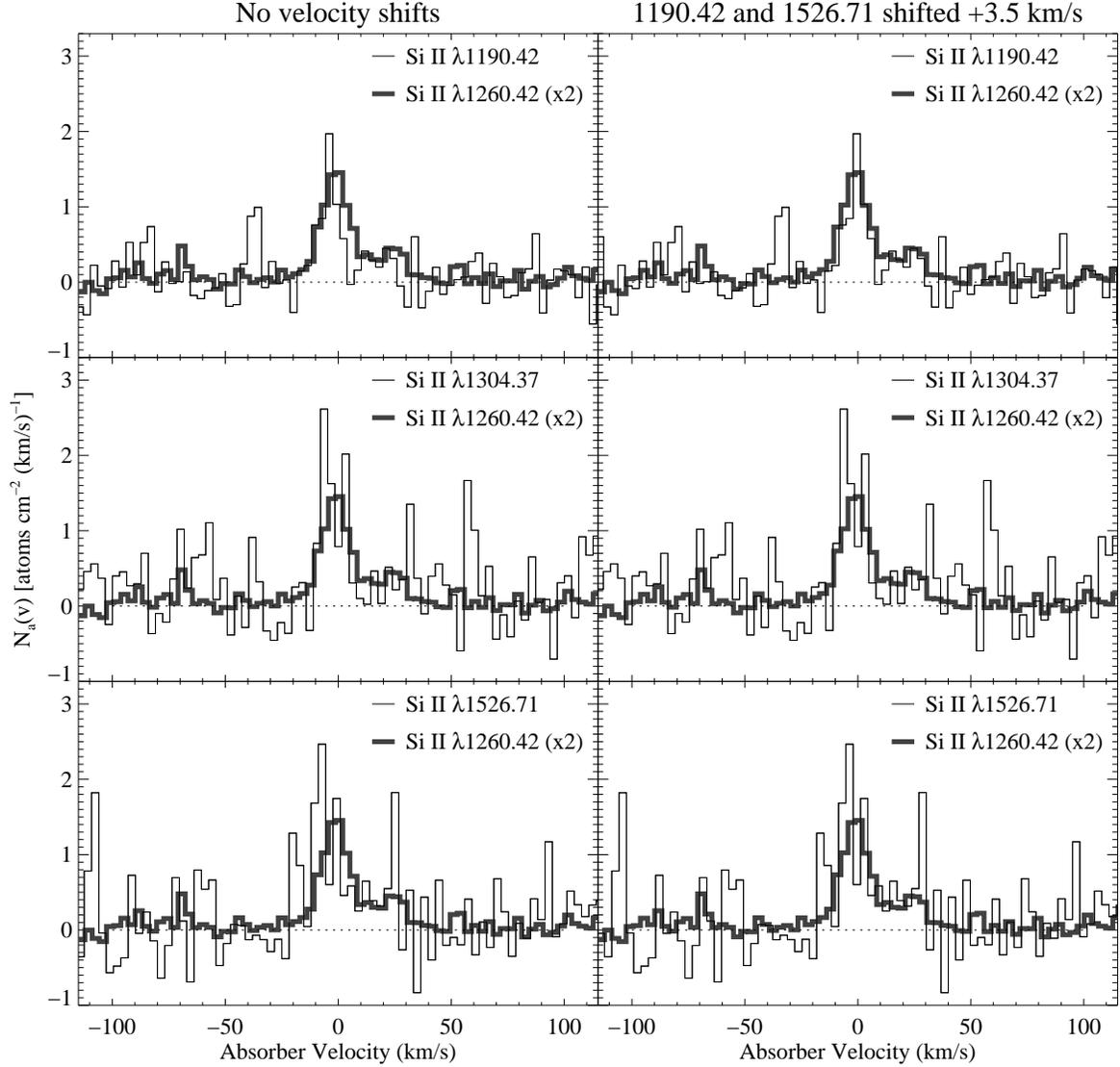}
\caption[]{Comparison of the apparent column density profile of
\ion{Si}{2} $\lambda$1260.42 to the \nav\ profiles of \ion{Si}{2}
$\lambda$1190.42 (top), \ion{Si}{2} $\lambda$1304.37 (middle), and
\ion{Si}{2} $\lambda$1526.71 (bottom). As in Figure~\ref{oshift}, the
left panels show the \nav\ profiles with no shifts applied to the
velocity scale. The \nav\ profiles in this figure are somewhat noisier
than the profiles shown in Figure~\ref{oshift}, but nevertheless,
close inspection reveals differences in the profile centroids. The
right panels show the same comparisons, but with the 1190.42 and
1526.71 \AA\ transitions shifted by +3.5 \kms , which appears to
provide better alignment of these lines.\label{sishift}}
\end{figure}

\begin{figure}
\plotone{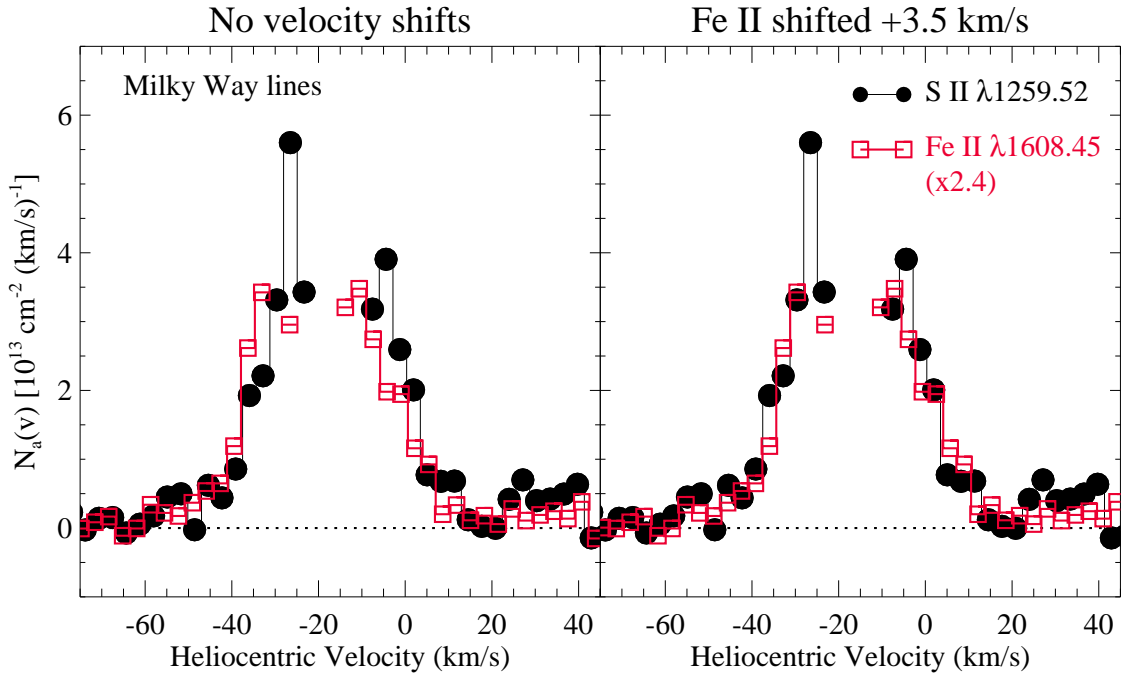}
\caption[]{The \nav\ profile of the Milky Way \ion{S}{2}
$\lambda$1259.52 transition (filled circles) vs. \nav\ of the Milky
Way \ion{Fe}{2} $\lambda$1608.45 line (open squares). In the left
panel, no shift has been applied to the wavelength scale, but in the
right panel the \ion{Fe}{2} line has been shifted by +3.5 \kms . The
profiles are not plotted in the line cores where both the \ion{S}{2}
and \ion{Fe}{2} lines are saturated. These lines, which have nothing
to do with the transitions from the \zabs\ = 0.00632 absorber shown in
Figures~\ref{oshift} - \ref{sishift}, also indicate that a $\sim$1
pixel correction is required for the wavelength scale of the
\ion{Fe}{2} line.\label{feshift}}
\end{figure}

According to the Cycle 13 STIS Handbook, the relative wavelength
calibration of STIS spectra recorded with the MAMA detectors is
accurate to 0.25-0.5 pixels across an exposure, and the absolute
wavelength calibration (i.e., the zero point) is good to 0.5$-$1.0
pixels (Kim Quijano et al. 2003). We have examined many spectra
obtained with the STIS echelle modes (which use the MAMA detectors),
and in general we have found the wavelength calibration to be
excellent. Jenkins \& Tripp (2001) noticed small wavelength
calibration errors in a particular order of a set of E230H spectra,
but in various projects using STIS E140M spectra of a large number of
targets, we have usually found that absorption lines (that are
appropriate for comparison) are well-aligned across the full
wavelength range of STIS echelle spectra, in agreement with the
accuracies reported in the STIS Handbook. Consequently, we were
surprised to find evidence of somewhat larger wavelength calibration
errors in the STIS E140M echelle spectrum of PG1216+069. We find
indications that the relative wavelength calibration errors are as
large as $\sim$1 pixel. Some of this evidence is shown in
Figures~\ref{oshift}-\ref{feshift}. Figures~\ref{oshift} and
\ref{sishift} compare \nav\ profiles of several of the absorption
lines in the \zabs\ = 0.00632 absorption system.  In the left panels
in these figures, no shifts have been applied to the wavelength scale
in the final reduced spectrum. Differences in the velocity centroids
of these lines are readily apparent. In the right panels in
Figures~\ref{oshift}-\ref{sishift}, shifts of $\sim$1 pixel have been
applied to several of the lines, as described in the
captions. Clearly, the $\sim$1 pixel corrections improve the alignment
of the various lines observed in this absorber. In this paper, we have
shifted lines at $\lambda _{\rm ob} \ <$ 1200 \AA\ and $\lambda _{\rm
ob} \ >$ 1330 \AA\ by +3.5 \kms .

Do we expect the transitions in Figures~\ref{oshift}-\ref{sishift} to
be well aligned? Could some multiphase/multicomponent aspect of the
absorber cause real differences in the kinematics of the various
species compared in these figures?  The \ion{Si}{2} transitions in
Figure~\ref{sishift} should all show the same kinematical
behavior. Multiple components that are blended and unresolved could
lead to confusing line profile differences in comparisons of strong
and weak transitions of a particular species (because column density
components can be detectable in the strong lines but not the weak
ones). However, comparisons of the weaker \ion{Si}{2} transitions (e.g.,
1304.37 vs. 1526.71) as well as weak vs. strong transitions (e.g.,
1260.42 and 1526.71) consistently require the same shifts, which
suggests that this not a component structure effect. Moreover, several
other line comparisions consistently indicate that the same wavelength
scale corrections are needed. We find that lines at $\lambda >$ 1330
\AA\ and $\lambda <$ 1200 \AA\ must be shifted compared to lines
within the 1200-1330 \AA\ range. For example, Figure~\ref{oshift}
shows several species that are the dominant ionization stage in
\ion{H}{1} regions; these lines require the same shifts seen in the
\ion{Si}{2} lines in Figure~\ref{sishift}. In principle, dominant ions
such as \ion{C}{2}, \ion{Si}{2}, and \ion{Fe}{2} could arise in
ionized gas as well as the neutral gas, and this could cause profile
differences, but the fact that the same shifts are indicated by the
data in Figure~\ref{oshift} argues that this is not an ionization
effect.  Similarly, Figure~\ref{feshift} overplots the Milky Way
\ion{S}{2} $\lambda$1259.52 and \ion{Fe}{2} $\lambda$1608.45 \nav\
profiles, and the same shift leads to significantly improved alignment
of these Galactic lines.  The \ion{S}{2} and \ion{Fe}{2} lines in
Figure~\ref{feshift} are expected to have very similar strengths based
on the abundances of these species in the ISM and their atomic data,
and these lines are expected to be well-aligned. In this case, greater
depletion of Fe by dust might generate profile differences, but since
the same velocity shift is again mandated, we believe that the
difference in the left panel of Figure~\ref{feshift} is more likely
due to wavelength calibration errors.

We have also reduced the PG1216+069 E140M spectrum using the STScI
pipeline, and we find that the final pipeline spectrum is very similar
to our spectrum (which was reduced with the STIS Team software) and
shows the same evidence of wavelength scale errors.  We conclude that
the PG1216+069 spectrum shows unusually large relative wavelength
errors compared to most STIS echelle spectra that we have studied.
However, even with these larger errors, the wavelength calibration is
still very good compared to most archival UV spectra obtained with
other {\it HST} instruments and other space-borne UV facilities.

\begin{deluxetable}{lcccc}
\tablewidth{0pc}
\tablecaption{\ion{H}{1} Column Densities: Virgo sub-DLA and Adjacent Lines\label{h1col}}
\tablehead{Observed & Line & \zabs & log $N$(\ion{H}{1}) & Method\tablenotemark{a} \\
Wavelength & Identification & \ & \ & \ }
\startdata
1215.60 & \ion{H}{1} Ly$\alpha$ & 0.0 & 20.17$\pm$0.03 & 1 \\
1220.23 & \ion{H}{1} Ly$\alpha$ & 0.00375 & 13.64$\pm$0.09 & 2 \\
1223.34 & \ion{H}{1} Ly$\alpha$ & 0.00632 & 19.32$\pm$0.03 & 1 \\
\enddata
\tablenotetext{a}{Column densities and uncertainties determined using the following methods: (1) fit to the Lorentzian wings of the damped Ly$\alpha$ profile as described in Jenkins et al. (1999), and (2) Voigt profile fit to the weak Ly$\alpha$ line.}
\end{deluxetable}

\begin{deluxetable}{lcccc}
\tablewidth{0pc}
\tablecaption{Metal Line Equivalent Widths and Integrated Column Densities\label{intprop}}
\tablehead{ \ \ \ Species \ \ \ & $\lambda _{0}$\tablenotemark{a} & log $f\lambda_{0}$\tablenotemark{a} & $W_{\rm r}$ & log $N_{\rm a}$ \\
  \  & (\AA ) & \ & (m\AA ) & \\ \hline
\multicolumn{5}{c}{All Components\tablenotemark{b}}}
\startdata
\ion{O}{1}\dotfill & 1302.17 & 1.830 & 94.1$\pm$8.3 & 14.32$\pm$0.05 \\
\ion{C}{2}\dotfill & 1334.53 & 2.229 & 110.0$\pm$8.6 & $>$13.96\tablenotemark{c} \\
\ion{Si}{2}\dotfill & 1190.42 & 2.543 & 66.8$\pm$15.2 & 13.40$^{+0.09}_{-0.12}$\tablenotemark{c} \\
     \              & 1260.42 & 3.174 & 140.0$\pm$7.8 & $>$13.15\tablenotemark{c} \\
     \              & 1304.37 & 2.078 & 43.8$\pm$9.1 & 13.60$\pm$0.10 \\
     \              & 1526.71 & 2.303 & 76.3$\pm$19.5 & 13.61$^{+0.10}_{-0.13}$ \\ \hline
\multicolumn{5}{c}{Strongest Component Only\tablenotemark{d}} \\ \hline
\ion{O}{1}\dotfill & 1302.17 & 1.830 & 75.5$\pm$5.5 & 14.21$\pm$0.05 \\
\ion{N}{1}\dotfill & 1199.55 & 2.193 & ($-23.5\pm 10.5$) & $<$13.28 \\
\ion{C}{2}\dotfill & 1334.53 & 2.229 & 77.7$\pm$5.7 & $>$13.84\tablenotemark{c} \\
\ion{Si}{2}\dotfill & 1190.42 & 2.543 & 47.3$\pm$9.6 & 13.25$^{+0.09}_{-0.12}$\tablenotemark{c} \\
     \              & 1260.42 & 3.174 & 98.2$\pm$5.1 & $>13.03$\tablenotemark{c} \\
     \              & 1304.37 & 2.078 & 26.2$\pm$6.2 & 13.40$^{+0.09}_{-0.11}$ \\
     \              & 1526.71 & 2.303 & 53.5$\pm$12.9 & 13.45$^{+0.10}_{-0.14}$ \\ 
\ion{Fe}{2}\dotfill & 1608.45 & 1.970 & 16.7$\pm$5.2 & 13.23$^{+0.12}_{-0.14}$ \\
\ion{C}{4}\dotfill & 1550.78\tablenotemark{e} & 2.169 & (8.3$\pm$15.4) & $<13.36$ \\
\ion{Si}{4}\dotfill & 1402.77\tablenotemark{e} & 2.554 & ($-22.4\pm$11.3) & $<12.88$ 
\enddata
\scriptsize
\tablenotetext{a}{Rest-frame wavelength ($\lambda _{0}$) and oscillator strength ($f$) from Morton (1991) or Morton (1999).}
\tablenotetext{b}{Rest-frame equivalent width [$W_{\rm r} = W_{\rm ob}/(1 + z)$] and apparent column density integrated across the two main components of the absorber at $v = 0$ and 28 \kms .}
\tablenotetext{c}{The column density from direct integration of this line is underestimated due to unresolved line saturation (see text, \S~\ref{abmeas}).}
\tablenotetext{d}{Equivalent widths and column densities integrated over the velocity range of the strongest component only, i.e., the component at $v = 0$ \kms\ in Figure~\ref{normprofs}. Equivalent widths listed in parentheses have less than $3\sigma$ significance; these lines are not reliably detected, and we derive the $3\sigma$ upper limit on the column density from the equivalent width uncertainty assuming the linear curve of growth is applicable.}
\tablenotetext{e}{The stronger \ion{C}{4} $\lambda$1548.20 and \ion{Si}{4} $\lambda$1393.76 lines cannot be measured due to blending with unrelated lines from other redshifts.  Both lines of the \ion{N}{5} $\lambda \lambda$1238.82, 1242.80 doublet are blended, so no constraints can be placed on this species at this redshift.}
\end{deluxetable}

\begin{deluxetable}{lcccc}
\tablewidth{0pc}
\tablecaption{Component Velocities, Doppler Parameters, and Column Densities from Voigt-Profile Fitting\label{voigtfit}}
\tablehead{ \ \ \ Species \ \ \ & Fitted Lines & $v$\tablenotemark{a} & $b$ & log $N$ \\
  \  & (\AA ) & (\kms ) & (\kms ) & \ }
\startdata
\ion{O}{1}\dotfill  & 1302.17 & 0$\pm$1  & 7$\pm$1 & 14.32$\pm$0.09 \\
       \            & \       & 28$\pm$2 & 9$\pm$3 & 13.66$\pm$0.09 \\
\ion{C}{2}\dotfill  & 1334.53 & $-4\pm$1  & 6$\pm$1 & 14.21$\pm$0.30 \\
       \            & \       & 28$\pm$1 & 8$\pm$2 & 13.36$\pm$0.08 \\
\ion{Si}{2}\dotfill & 1190.42, 1260.42, 1304.37, 1526.71 & $-2\pm$1 & 6$\pm$1 & 13.44$\pm$0.06 \\
       \            & \       & 23$\pm$1 & 10$\pm$2 & 12.66$\pm$0.06 \\
\ion{Fe}{2}\dotfill & 1608.45 & $-4\pm$1  & 3$\pm$1 & 13.43$\pm$0.11 
\enddata
\tablenotetext{a}{Velocities in this table have been corrected for the wavelength calibration errors discussed in the appendix.}
\end{deluxetable}

\newpage

\clearpage

\clearpage

\clearpage

\end{document}